\RequirePackage{rotating}
\documentclass[fleqn,usenatbib]{mnras}

\usepackage{newtxtext,newtxmath}

\usepackage[T1]{fontenc}

\DeclareRobustCommand{\VAN}[3]{#2}
\let\VANthebibliography\thebibliography
\def\thebibliography{\DeclareRobustCommand{\VAN}[3]{##3}\VANthebibliography}

\usepackage{graphicx}	
\usepackage{amsmath}	
\usepackage{caption}
\usepackage{multirow}
\usepackage{multicol}        
\usepackage{threeparttable}
\usepackage{longtable}
\usepackage{nicefrac}
\usepackage{subfig}
\usepackage{stfloats}
\usepackage [english]{babel}
\usepackage [autostyle, english = american]{csquotes}
\MakeOuterQuote{"}

\makeatletter
\providecommand\phantomcaption{\caption@refstepcounter\@captype}
\makeatother

\newcommand\hst{{\it HST \/}}
\newcommand\HST{{\it Hubble Space Telescope \/}}
\newcommand\Mpeak{{$M_{peak}$ \/}}
\def\simgt{\lower.5ex\hbox{$\; \buildrel > \over \sim \;$}}
\def\simlt{\lower.5ex\hbox{$\; \buildrel < \over \sim \;$}}

\graphicspath{{./}{Figures/}}

\title[Novae in M51]{Novae in M51: a New, Much Higher Rate from Multi-epoch HST Data}

\author[Mandel et al.]{
Shifra Mandel,$^{1}$\thanks{E-mail: ss5018@columbia.edu}
Michael M. Shara,$^{2}$
David Zurek$^{2}$
Charlie Conroy$^{3}$
and Pieter van Dokkum$^{4}$
\\
$^{1}$Columbia Astrophysics Laboratory, Columbia University, New York, NY 10027, USA\\
$^{2}$Department of Astrophysics, American Museum of Natural History, New York, NY, USA\\
$^{3}$Department of Astronomy, Harvard University, Cambridge, MA, 02138, USA\\
$^{4}$Department of Astronomy, Yale University, New Haven, CT, 06511, USA
}

\date{Accepted 2022 September 27. Received 2022 September 26; in original form 2022 August 24}

\pubyear{2022}

\begin{document}
\label{firstpage}
\pagerange{\pageref{firstpage}--\pageref{lastpage}}
\maketitle

\begin{abstract}
Accurate determination of the rates of nova eruptions in different kinds of galaxies give us strong constraints on those galaxies' underlying white dwarf and binary populations, and those stars' spatial distributions.  Until 2016, limitations inherent in ground-based surveys of external galaxies -- and dust extinction in the Milky Way -- significantly hampered the determination of those rates and how much they differ between different types of galaxies.  Infrared Galactic surveys and dense cadence \HST (\textit{HST})-based surveys are overcoming these limitations, leading to sharply increased nova-in-galaxy rates relative to those previously claimed.  Here we present 14 nova candidates that were serendipitously observed during a year-long \textit{HST} survey of 
the massive spiral galaxy M51 (the "Whirlpool Galaxy").  We use simulations based on observed nova light curves to model the incompleteness of the \textit{HST} survey in unprecedented detail, determining a nova detection efficiency $\epsilon = 20.3$~percent.  
The survey's M51 area coverage, combined with $\epsilon$,  
indicates a conservative M51 nova rate of $172^{+46}_{-37}$~novae yr$^{-1}$, corresponding to a luminosity-specific nova rate (LSNR) of $\sim10.4^{+2.8}_{-2.2}$~novae yr$^{-1}$/$10^{10} L_{\odot,K}$.  Both these rates are approximately 
an order of magnitude higher than those estimated 
by ground-based studies, 
contradicting
claims of universal low nova rates in all types of galaxies determined by low cadence, ground-based surveys.  They demonstrate that, contrary to theoretical models, the \textit{HST}-determined LSNR in a giant elliptical galaxy (M87) and a giant spiral galaxy (M51) likely do {\it not} differ by an order of magnitude or more, and may in fact be quite similar. 
\end{abstract}



\section{Introduction} \label{sec:intro}

All cataclysmic variables (CVs) are binaries containing a white dwarf (WD) which accretes matter from a close companion.
A nova eruption is a bright (up to $10^6$ $L_{\odot}$) outburst that occurs when the envelope accreted onto the WD surface ignites in a thermonuclear runaway.  Nova characteristics (such as the recurrence rate, peak luminosity, and decay time) encode information about the WD and donor star, 
as well as the binary mass transfer rate during the millenia between nova eruptions \citep{Hillman2016,Hillman2020}.  Novae are our only means of studying CV populations (and indeed most binary populations) in galaxies beyond the Local Group.  In addition, the most rapidly accreting WDs in nova binaries can be progenitors of "standard candle" type Ia supernovae (SNIa) \citep{Hillman2016}, so these stars' importance extends beyond the domain of stellar evolution to cosmology.  

Given that CVs with high accretion rates from sub-giant companions and/or very massive WDs are likely SNIa progenitors \citep{Hillman2016}, the dependence of CV populations on the underlying stellar populations and the environments of their host galaxies is of great importance for determining whether SNIa are reliable standard candles.  Varying CV populations in different galaxy types could be an indicator of differing SNIa progenitor channels, with important implications for the determination of the Hubble constant $H_0$ using SNIa as distance indicators.  Differences in CV populations could also hint at different binary fractions and/or stellar evolution pathways in different types of galaxies.  

Despite their importance in understanding and testing models of binary stellar evolution, and implications for cosmological standard candles, a lack of consensus on the actual nova rates in galaxies has persisted for two decades.  On the basis of multiple ground-based surveys, \citet{Shafter2000} and \citet{Shafter2014} claimed that the luminosity specific nova rates (LSNR, i.e.  annual rate of novae per unit K-band luminosity) in different galaxy types are all similar, in the range of 1--3 novae yr$^{-1}$/$10^{10}L_{\odot,K}$.  
In contrast, the population synthesis studies of \citet{Matteucci2003}, \citet{Claeys2014} and \citet{Chen2016} suggested that order-of-magnitude differences in nova rates and LSNR should exist between elliptical, spiral, and especially starburst galaxies.  This is because rapid and massive star formation should produce a plethora of mass-transferring binaries containing high-mass WDs in spiral and starburst galaxies.  Novae which erupt on high-mass WDs do so after accreting relatively low-mass envelopes \citep{Yaron2005}.  Such novae can thus erupt more frequently than those associated with low mass WDs, so that the LSNR in spiral and especially in starburst galaxies are predicted to greatly exceed the corresponding rates in elliptical galaxies.

Using a \HST (\textit{HST}) survey of the massive elliptical galaxy M87, \citet{Shara2016} showed that ground-based surveys of external galaxies fail to detect fainter novae and/or those with short decline times and/or those near the bright centers of galaxies.  These effects cause ground-based surveys to systematically underestimate the true nova rates in galaxies.  In the case of M87, these effects led to the ground-based underestimate of the M87 nova rate by a factor of 2-4; the \hst- determined LSNR in the K-band was shown to be $7.88_{-2.6}^{+2.3}$ novae yr$^{-1}$/$10^{10}L_\odot,_{K}$ \citep{Shara2016}.  Confirmation of ground-based nova rate underestimates was provided by \citet{Mroz2016}, who demonstrated that the LSNR in the Large Magellanic Cloud (LMC) is much higher than previous ground-based estimates, and is comparable to the M87 LSNR.  This finding was further supported by \citet{De2021}, who discovered a sizable population of Galactic novae (in the infrared) that have gone undetected in over a century of optical searches, and \citet{Kawash2021}, who found that approximately half of all Galactic novae are hidden from current surveys by extinction.  

These discoveries (of much higher than previously claimed LSNR) in a giant elliptical (M87), a barred spiral (the Galaxy), and a dwarf irregular galaxy (LMC), were carried out via surveys with much longer baselines, denser time coverage and/or deeper magnitude limits than all previous surveys.  They argue strongly against the claim that the LSNR is relatively low in all galaxies, as the earlier, shallower and sparser cadence coverage of \citet{Shafter2000} and \citet{Shafter2014} suggested.  They highlight the need for deep, unbiased surveys of other types of galaxies to confirm that the LSNRs are much higher than previously thought, and to test whether they vary with galaxy type, as predicted by binary population synthesis models.  In particular, a giant Sc-type spiral galaxy has not yet been so studied.  \hst is especially suited to such investigations because of its unparalleled angular resolution and consequent sensitivity, its very small and virtually constant point-spread function, its insensitivity to lunar phase and its immunity to atmospheric seeing.

The massive Sc-type spiral galaxy M51 (the Whirlpool Galaxy, NGC 5194) has been surveyed for novae only once, over 20 years ago \citep{Shafter2000}.  Narrowband $H\alpha$ and broadband $R$ images centered on M51, and covering $16'x16'$ (the entire galaxy), were taken with the Kitt Peak National Observatory 4-meter telescope at four well-separated epochs in 1994 and 1995.  These data led to the discovery of nine novae.  Allowing for gaps in coverage and other sources of incompleteness (such as a limiting absolute magnitude detection limit of -7.7 $\pm$ 0.22), \citet{Shafter2000} derived a rate of 18 $\pm$ 7 novae yr$^{-1}$ in M51.  This corresponds to an M51 LSNR of 1.09 $\pm$ 0.47 novae yr$^{-1}$/$10^{10}L_{\odot,K}$.  

M51 was the subject of an \hst observing campaign that began in 2016 and continued for nearly a year \citep{Conroy2018}.  The stated goal of that survey was to catalog and categorize all luminous stellar variables within a significant fraction of that galaxy.  Given its optimal orientation on the sky -- M51 is face-on, which minimizes internal reddening, and benefits from a fairly low Galactic extinction that is approximately constant across the field -- this survey also offers an excellent opportunity for the most complete and unbiased study of the novae in a massive Sc-type spiral galaxy to date.  Although the irregular cadence and sometimes weeks-long gaps between observations of M51 are not ideal for a nova survey, the proximity of M51 (relative to M87) allows us to detect novae in \hst images even if they are intrinsically faint (hence much too faint to detect from the ground) and/or after they have dimmed several magnitudes from maximum light.  In particular, the archival \hst imaging dataset of M51 enables just the second, head-to-head comparison (after M87) of ground-based versus \hst- derived nova rates in the same galaxy.

Section \ref{sec:data} describes the data collected during the M51 \hst observing campaign.  Section \ref{sec:novae} describes our search for and identification of nova candidates 
and their properties.  
In Section \ref{sec:efficiency} and in section \ref{sec:realsims} we describe the details of the simulations we conducted to investigate, respectively, the detection efficiency and the incompleteness of our \hst survey.   In Section \ref{sec:discussion} we discuss our findings and their implications, and we summarize our results in Section \ref{sec:conclusion}.

\section{\hst Imaging Data} \label{sec:data}

\begin{figure}
\includegraphics[width=\linewidth]{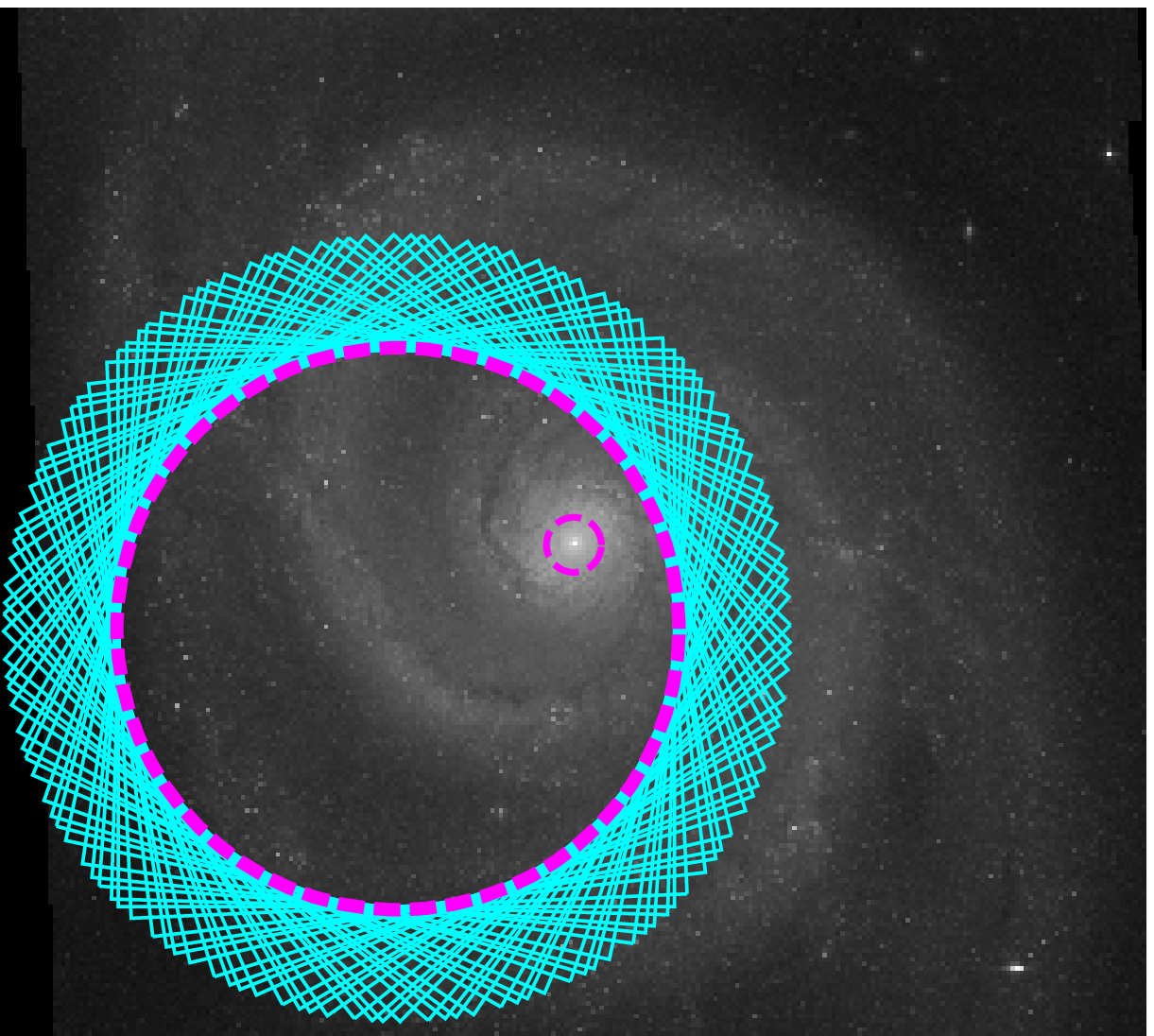}
\caption{\hst ACS $I_{814}$-band mosaic of M51.  The footprints of the 34 observations used for this study are shown in cyan; the large dashed magenta circle marks the region consistently observed and included in the photometric data.   \label{fig:hst_f814w_fpr}}
\end{figure}

The \hst observing campaign of M51 was conducted over the course of 345 days using the Advanced Camera for Surveys (ACS) Wide-Field Channel (WFC) $F606W$ and $F814W$ filters (hereafter $V_{606}$ and $I_{814}$, respectively).  Each of the 34 imaging epochs, with a total 2.2 ksec exposure in each of the $V_{606}$ and $I_{814}$ bands, covered $\sim 40$~percent of the galaxy's $I_{814}$ flux (see Section \ref{subsec:deteff}).  While the {\it average} gap between observations was $\sim10$~days, the survey cadence ranged from 4 to 24 days.  Figure \ref{fig:hst_f814w_fpr} shows the \hst fields of view (FOV) of those 34 epochs.  Note that the FOVs rotate to maintain optimal pointing of \hst's solar panels throughout the course of the year.  We considered only the portion of the FOVs that was covered in every epoch (see Figure \ref{fig:footprint} for more details).  

To create the M51 star catalog, point-spread-function (PSF) photometry was performed using the \textsc{DOLPHOT} software package \citep{Dolphin2000, Conroy2018}.  The regions around six bright foreground stars were masked to avoid contamination.  The central 10 arcsec were also excluded, because the region is extremely crowded, making PSF photometry impracticable.  A detailed description of the observations and the methods used to extract the photometric data are given in \citet{Conroy2018}.  In total, the \hst data yielded photometric measurements at 34 epochs in two passbands for $\sim 1.39$~million stars.  

The survey's magnitudes are on the Vega zero point (\textit{VEGAMAG}) system.  All absolute magnitudes were computed assuming an M51 distance modulus of 29.67 with Galactic extinctions in the direction of M51 of $A_{606} = 0.086$ and $A_{814} = 0.053$~mag \citep{McQuinn2016}.

\section{M51 Nova Search and identification} \label{sec:novae}

\setlength{\tabcolsep}{18pt}
\begin{center}
\begin{table*}
\caption{Nova Candidates in M51 \label{tab:nclist}}  
\begin{center}
\begin{tabular}{c c c c c c c} 
 \hline\hline
 Nova & M51 Nucleus Offset & RA (J2000) & DEC (J2000) & M$_{peak}$ & M$_{peak}$ & t$_{peak, V}$ \\ 
  & [arcsec] & [deg] & [deg] & ($V_{606}$) &  $(I_{814})$ & [MJD] \\ [1ex] 
 \hline
 1  & 74.31  & 202.49935  & 47.19936  & -8.51  & -8.97  &  57675 \\
2  & 10.77  & 202.46526  & 47.19466  & -7.99  & -8.28  &  57936  \\
3  & 23.36  & 202.46145  & 47.19867  & -7.79  & -8.27  &  57971  \\
4  & 114.87  & 202.48422  & 47.16494  & -7.78  & -8.04  &  57858  \\
5  & 15.55  & 202.47586  & 47.19589  & -7.73  & -8.26  &  57946  \\
6  & 45.85  & 202.48117  & 47.18525  & -7.26  & -7.68  &  57971  \\
7  & 30.26  & 202.45930  & 47.19994  & -7.07  & -7.33  &  57666  \\
8  & 33.21  & 202.47781  & 47.18792  & -6.97  & -7.42  &  57666  \\
9  & 119.21  & 202.51632  & 47.20461  & -6.89  & -7.22  &  57760  \\
10  & 54.92  & 202.48981  & 47.20186  & -6.72  & -7.34  &  57993  \\
11  & 47.62  & 202.48200  & 47.20544  & -6.41  & -7.41  &  57760  \\
12  & 67.81  & 202.49248  & 47.18465  & -6.17  & -6.48  &  57699  \\
13  & 94.0  & 202.47425  & 47.16934  & -6.03  & -6.91  &  57925  \\
14  & 22.89  & 202.47877  & 47.19641  & -5.53  & -6.50  &  57666  \\
\hline 
\end{tabular} \\ [3 pt]
Fourteen nova candidates discovered in the \hst observations of M51, listed in order of observed peak luminosity.  
\end{center} 
\end{table*}
\end{center}

The four defining characteristics we used to identify potential M51 nova candidates (cf.  \citet{Shara2016}) among the $\sim 1.39$~million stars in the \hst dataset are:

\begin{itemize}
\item a peak absolute magnitude brighter than -5 in the $V_{606}$ and $I_{814}$ bands, 
\item a decrease from peak brightness of at least 2 mag over the duration of the observing campaign, 
\item a "blue" color (average near maximum light) of $V_{606} - I_{814} <0.50$~mag, and
\item no apparent periodic variability.
\end{itemize}

These criteria are satisfied by virtually all known novae, and are deliberately over-conservative so as not to miss any reasonably identifiable candidates on a first pass.  Approximately 1,000 preliminary nova candidates were selected using the above parameter and light curve constraints.  Every candidate's light curve was visually inspected, allowing us to weed out eclipsing, periodic, and spurious sources.  [A number of discarded sources could have been either novae or other highly variable (and similarly "blue"-colored) stars, like cepheids or luminous blue variables (LBVs).  We retained only sources that showed the canonical single-peaked fast rise and exponential decline that is most often observed in novae, but not in other highly variable stars, although nova light curves can take a variety of shapes \citep{Strope2010}.]  All surviving candidates 
were visually inspected in the \hst images.  Fourteen final nova candidates remained.

Table \ref{tab:nclist} lists the 14 nova candidates we identified in the \hst dataset, including their angular offsets from M51's nucleus, J2000 coordinates, absolute magnitudes ($V_{606}$- and $I_{814}$-band) at maximum light, and the date of observed maximum brightness.  (Note that the actual peak luminosity for these nova outbursts could have been higher by up to several magnitudes; because of the gaps between \hst observations, most were likely detected during their decline.)  We deliberately  
omit $t_2$ because, due to the large gaps between the \hst observations of M51, the uncertainties in $t_{peak}$ and $M_{peak}$ -- both of which are required to evaluate $t_2$ -- are too great to allow for meaningful estimates of the decline time.

\begin{figure*}
\includegraphics[width=\textwidth]{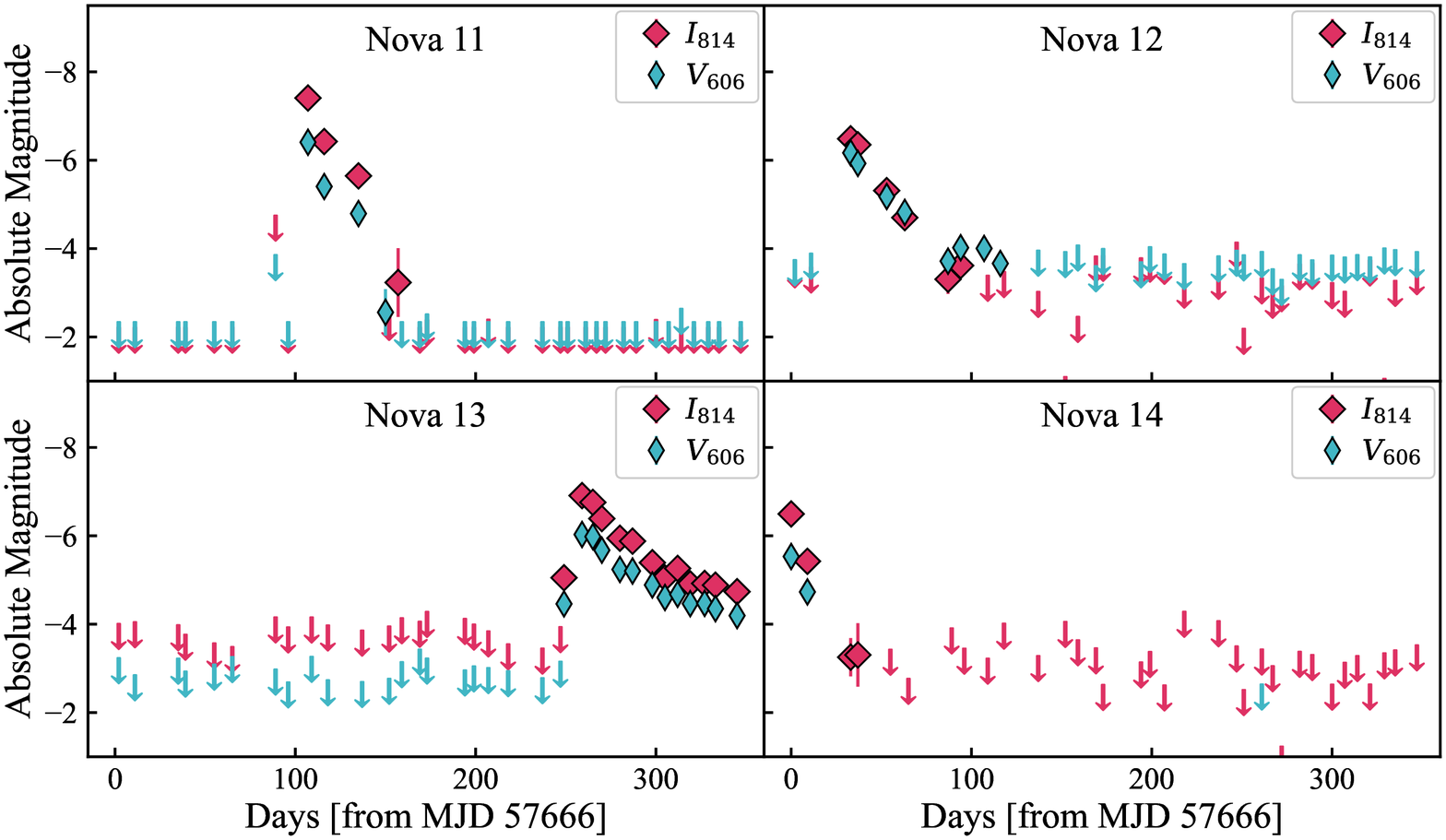}
\caption{$V_{606}$- and $I_{814}$- band light curves (shown in blue and red, respectively) for the 14 nova candidates in M51.  Arrows mark upper limits.  Where no error bars are visible, magnitude errors are smaller than the data point markers.  \label{fig:lc14}}
\end{figure*}

\begin{figure*}
\centering
\includegraphics[width=0.75\textwidth]{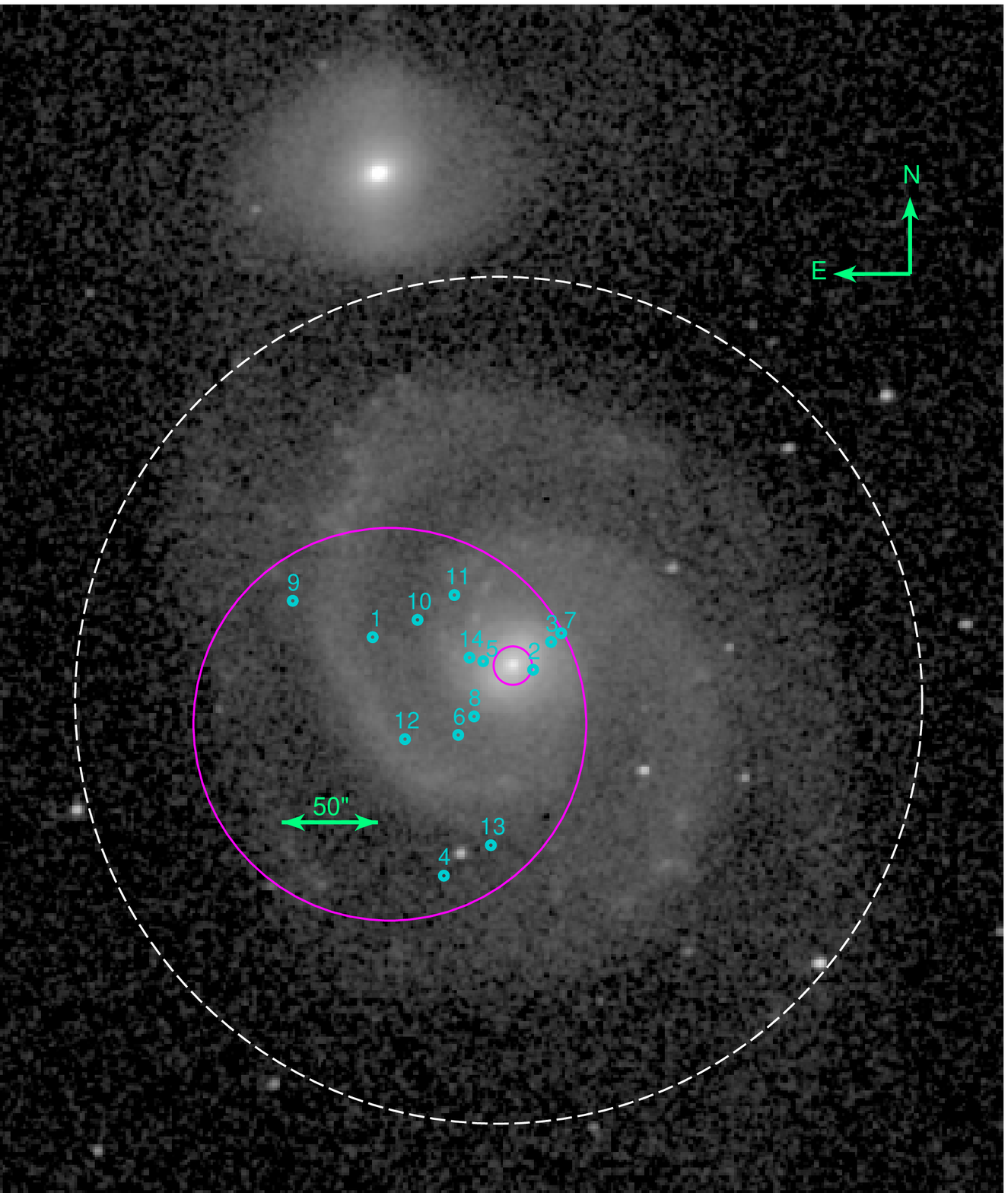}
\caption{2MASS K-band image showing the approximate "footprint" (cf.  Figure \ref{fig:hst_f814w_fpr}) of the \hst survey (magenta), 
an area of radius $\sim102$~arcsec.  
The central $10$~arcsec of M51 were excluded from our photometric analysis.  
The M51 light fraction covered in this survey - $40$~percent - was computed with the assumption that the light inside the white circle (radius 220 arcsec) constitutes all the light from M51 (see \ref{subsec:deteff} for more).  The locations and IDs of the nova candidates we identified are shown in turquoise.  \label{fig:footprint}}
\end{figure*}

The light curves for the 14 novae are shown in Figure \ref{fig:lc14}.  
"Postage stamp" difference images of all novae in Table \ref{tab:nclist} for each observing epoch are shown  
in Appendix \ref{sec:novim}.  The spatial distribution of the novae can be seen in Figure \ref{fig:footprint}, in which the positions of the 14 novae are overlaid on an image of M51.  \\

\begin{figure*}
\includegraphics[width=\textwidth]{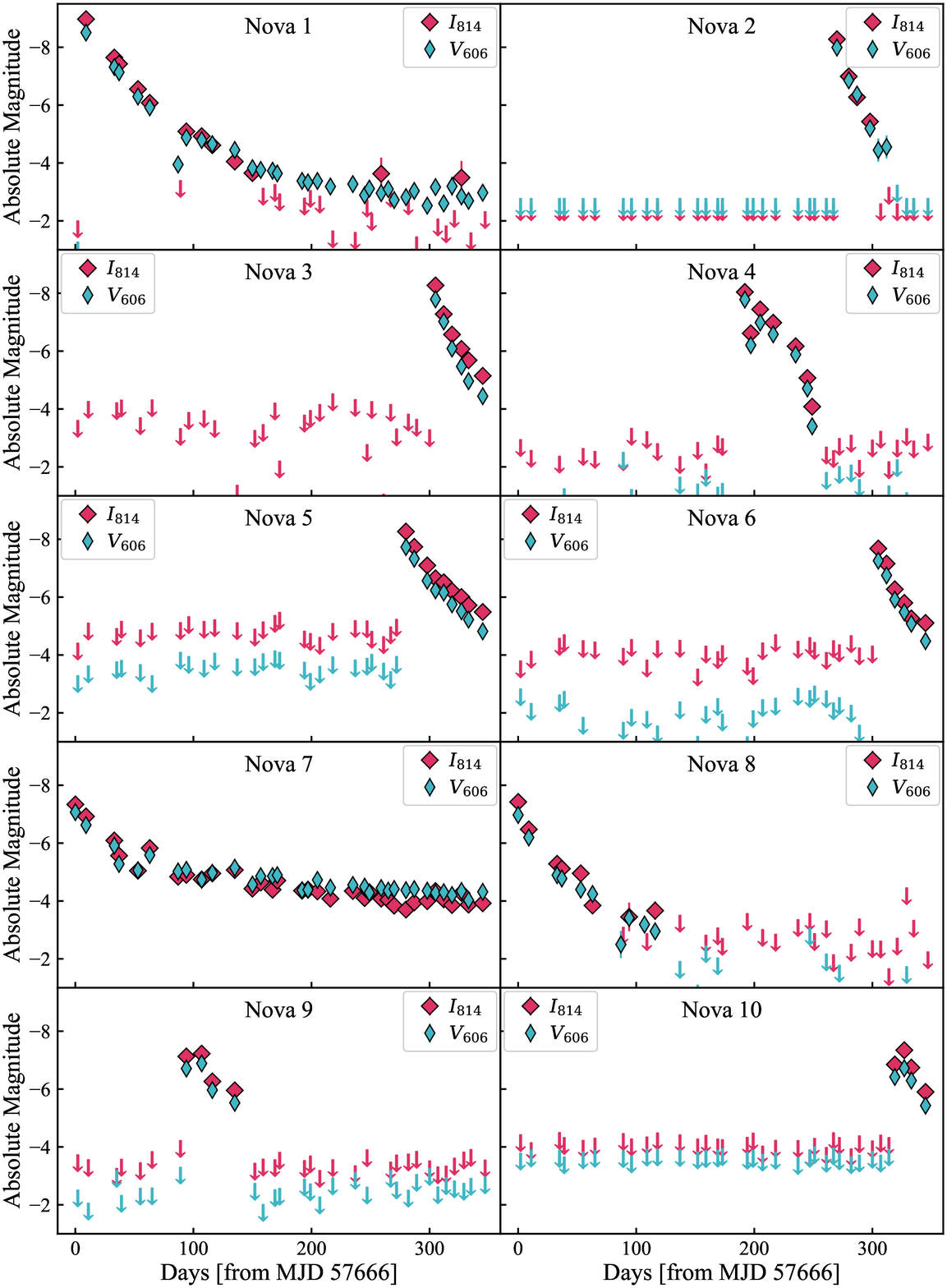}
\end{figure*}

\section{M51 Survey Nova Detection Efficiency} \label{sec:efficiency}

In addition to observing cadence, the observational properties that have the largest effect on whether a nova outburst will be detected or missed are that nova's peak luminosity ($M_{peak}$) and the decline time.  The latter is often defined as $t_2$, which is the time required for the luminosity to decrease from $M_{peak}$ by two magnitudes.  The non-uniform cadence of -- and large gaps between -- the \hst observations of M51 must necessarily hamper the detection of rapidly fading novae and the peak luminosities $L_{peak}$ of most outbursts.
To quantify the \emph{detection efficiency} ($\epsilon$) of our survey as a function of $M_{peak}$ and $t_2$ (as opposed to its \emph{incompleteness}, which we address in Section \ref{sec:realsims}), we generated an idealized set of artificial novae covering essentially the entire ranges of observed nova $M_{peak}$ and $t_2$.  (Not all combinations of nova $M_{peak}$ and $t_2$ are observed in nature, particularly the combination of most luminous $M_{peak}$  {\it and} longest $t_2$, but their inclusion is nonetheless instructive).  

We adopted $M_{peak}$ values of $[-10, -9.75, -9.5, \cdots, -5]$~mag and $t_2$ values of $[2, 5, 8, \cdots, 152]$~days.  All combinations of these 21 values of $M_{peak}$ and 51 values of $t_2$ produced a total of 1,071 artificial nova light curves.  Each artificial nova was assumed to decline exponentially, beginning at $M_{peak}$ and fading through the \hst detection threshold for the M51 survey, $V_{606} \approx 27.5$~mag.  
Each synthetic nova outburst was then begun at day ($t_0$) $[0, 1, 2, \cdots, 345]$ (corresponding to the length of the \hst observing campaign of M51) and sampled with the \hst observing cadence.  This range of outburst start times was necessary to account for biases in detectability based on the \hst observing cadence near the eruption date, as the gap between \hst epochs varied between 4 and 24 days throughout the campaign.  

\begin{figure*}
\begin{center}
\includegraphics[width=\textwidth]{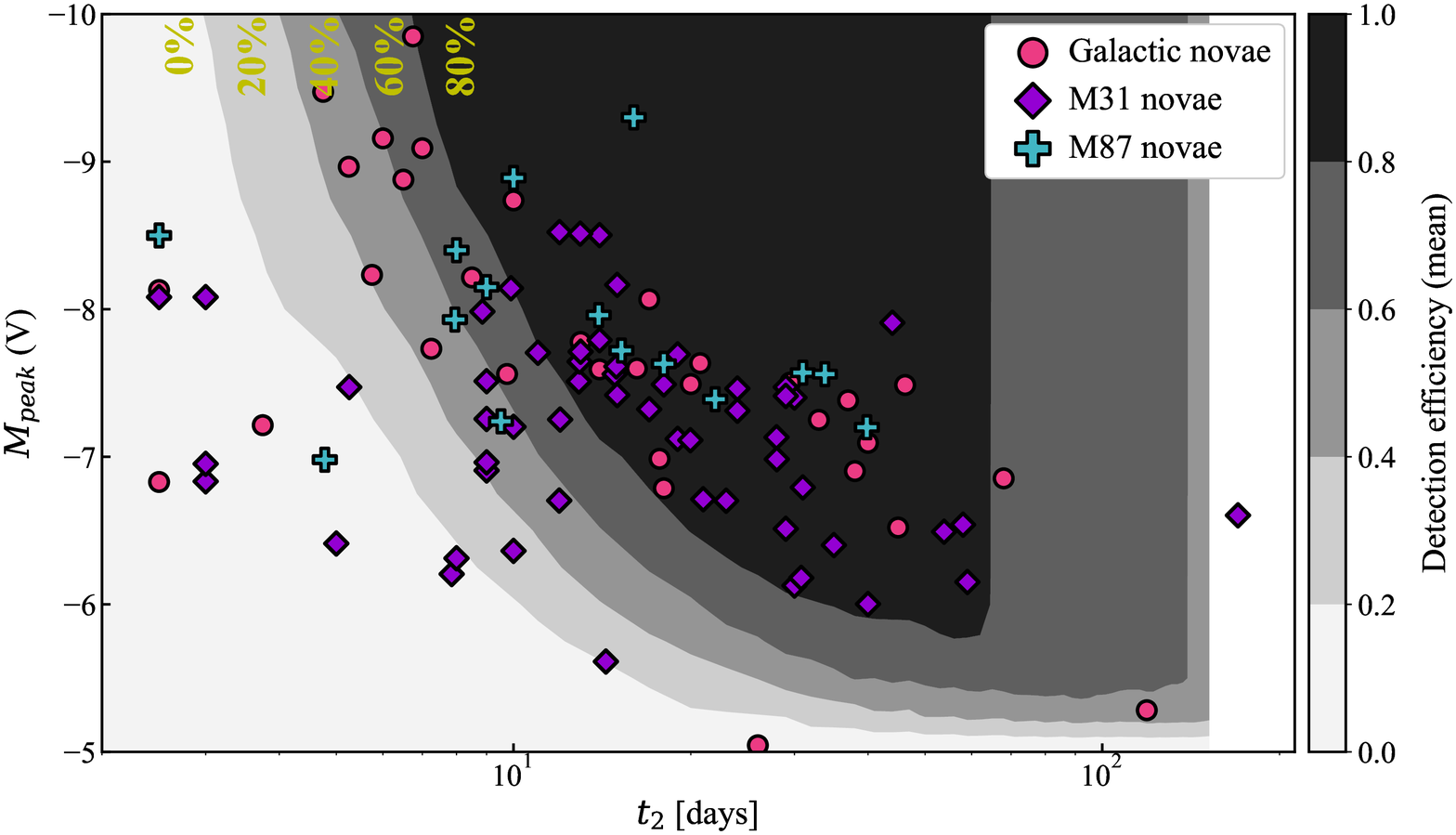}
\caption{Detection efficiency of synthetic novae for the full range of nova \Mpeak and $t_2$ discussed in Section \ref{sec:efficiency}, sampled at the M51 \hst observing cadence.  Contours denote a 20 percent change in efficiency.  Galactic, M31, and M87 novae are overlaid in pink, violet, and blue, respectively.
\label{fig:grideff}}
\end{center}
\end{figure*}

This process yielded $346\times1071 = 370,566$~light curves, sampled with the \hst observing cadence.  Each light curve was evaluated to determine whether the corresponding nova would have been detected as such via our $V_{606}$-band selection parameters, listed in Section \ref{sec:novae}.  This yields the survey's detection efficiency 
as a function of $M_{peak}$ and $t_2$.  The results are plotted in Figure \ref{fig:grideff}.  As expected, the figure demonstrates that novae with brighter $M_{peak}$ and longer $t_2$ were more likely to be detected than their fainter/faster counterparts, except when their decline times ($t_2$) stretched to months.  The latter effect is due to the decrease in observed variability when the decline time is very long.  

The existence of intrinsically faint novae with small $t_2$ (the so-called "faint/fast novae") was predicted in the 2005 compendium of nova models with a wide range of WD masses and accretion rates \citep{Yaron2005}.  They were first detected observationally by \citet{Kasliwal2011} in M31, then by \citet{Shara2016} in M87, and are now understood to be common.  Their ubiquity and importance in determining our survey's incompleteness are apparent in Figure \ref{fig:grideff}, where we overlaid the best optical samples of Galactic, M31, and M87 novae that are currently available on the detection efficiency plot just described.  

If they erupted in M51, a significant fraction of all novae observed in M87, M31 and the Galaxy would 
nearly always be missed in the current survey because they display $t_2<10$~days.  
The poor detectability of novae with $t_2 \simlt10$~days in our survey is a consequence of the large gaps between the \hst observations of M51.  
Many novae are observed to erupt with $t_2 \simlt10$~days, which, as noted above, 
are very difficult to detect in this survey.  Conversely, almost all novae with  $t_2>20$~days and $M_{peak}\simlt-8$ would be detectable (assuming a simplified exponential decline shape\footnote{The full range of nova light curve morphologies, including some that are considerably more complex than the simplistic models we adopt here, is detailed in \citet{Strope2010}.  We use those and other realistic light curves in section \ref{sec:realsims} below when we determine our survey's incompleteness.}).  No such novae have ever been reported, or predicted by large suites of nova models \citep{Yaron2005}.  The lack of detection of any such novae in our M51 \hst survey (despite the excellent detection efficiency associated with them) is further evidence that such objects are very rare, or do not exist.

\section{Incompleteness Simulations using Observed Novae} \label{sec:realsims}

\begin{figure*}
\includegraphics[width=1.\textwidth]{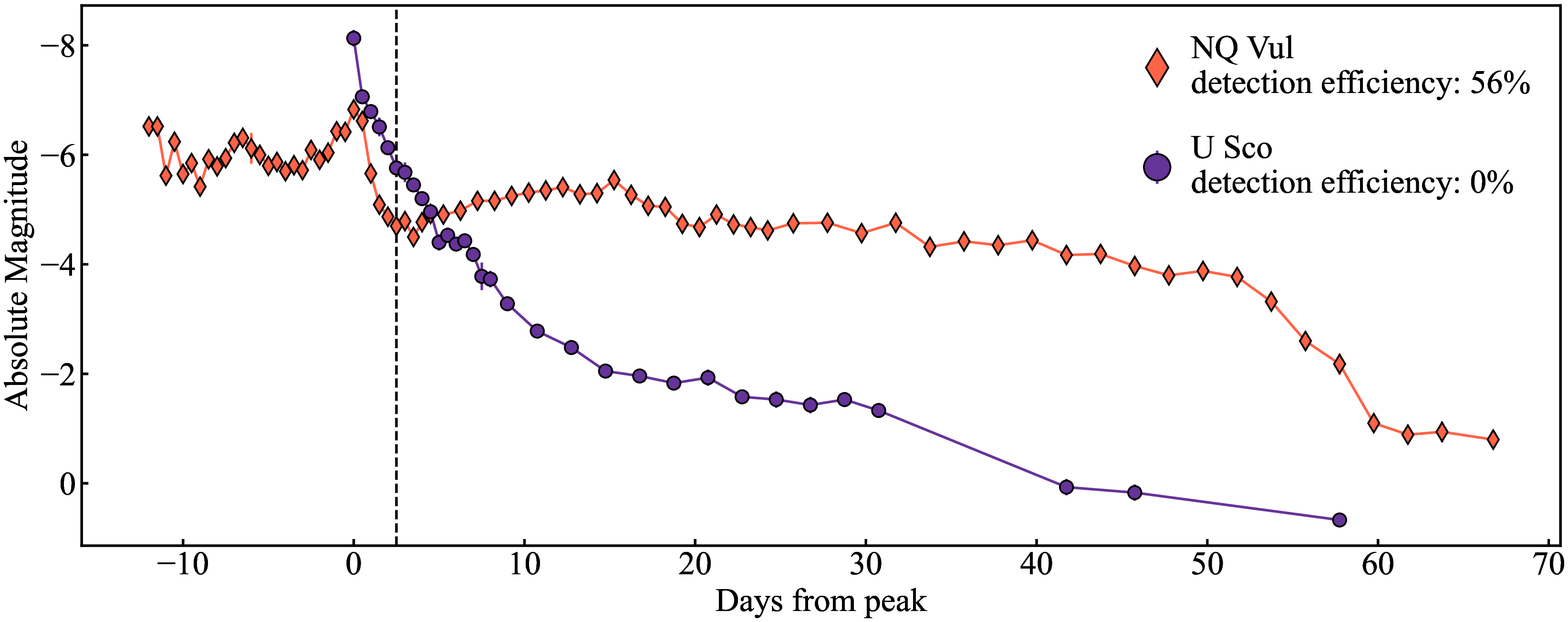}
\caption{Light curves of two Galactic novae, NQ Vul (orange) and U Sco (purple).  These novae have identical $t_2$ (dotted line) but different decline shapes, resulting in a lower detection efficiency in our M51 survey for the \textit{brighter} nova (U Sco).  
This demonstrates the importance of modelling detection efficiency using real nova light curves, as \Mpeak and $t_2$ alone are not reliable predictors for detectability.
\label{fig:lcvar}}
\end{figure*}

To determine the 
nova rate in M51,
we must first measure our survey's incompleteness viz. the fraction of novae that erupted in M51 during our survey but which were not detected.
While the idealized, highly simplified simulations described in Section \ref{sec:efficiency} help us understand how luminosity and decline time affect the detectability of novae subject to the \hst M51 survey cadence, the reality is more complex.  This is because the \emph{shape} of a nova light curve, which is generally \emph{not} exponential throughout, also plays a crucial role in its detectability, as demonstrated in Figure \ref{fig:lcvar}.  The decline of a nova's brightness is often fastest immediately following the outburst peak, though the opposite sometimes occurs.  Some novae reach a steady brightness plateau for days or weeks during their declines, while others undergo deep dips as dust forms in their ejecta.  The rates of change in luminosity, and when those changes occur, vary greatly among well-sampled Galactic novae \citep{Strope2010}.  Thus the irregularly spaced epochs of this survey must be convolved with a set of realistic light curves, representative of M51 novae, to determine our survey's incompleteness.  

There is no published set of well-sampled light curves of novae in M51 or in any other Sc-type galaxy.  
Thus we have produced a series of simulations using the three most realistic and complete sets of light curves available from well-observed novae in our Galaxy, M31 and M87.  A list of all novae that were used in the following simulations of \ref{subsec:gal} - \ref{subsec:m87} are found in Table \ref{simnovae} in Appendix \ref{sec:novlists}.  A summary of the results 
is shown in table \ref{tab:simeff}. 

\subsection{Simulation I.  Galactic Novae} \label{subsec:gal}

The largest collection of well-sampled novae starting near maximum brightness and extending well into the decline phase comprises, not surprisingly, Galactic novae.  \citet{Strope2010} published the light curves of 93 Galactic novae, most of which were well monitored from outburst peak through several magnitudes of decline.  Of these, 32 have well-determined distances that allow us to calculate their absolute magnitudes, either from \textit{Gaia} parallaxes \citep{Schaefer2018} or through the blackbody flux of their giant companions \citep{Ozdonmez2018}.  

This largest Galactic nova sample with well-calibrated luminosities is unfortunately far from unbiased.  In particular, it is conspicuously lacking in faint/fast novae \citep{Kasliwal2011, Shara2016}, which were missed throughout 20th century nova searches.  As Figure \ref{fig:grideff} demonstrates, our M51 survey's detection efficiency $\epsilon$ for novae with $t_2 \simlt10$ days is close to zero.  While we investigated whether each of these 32 Galactic novae would be identifiable as such \emph{had it occurred within M51 during our \hst observing campaign}, the lack of faint/fast novae in the Galactic sample means that its incompleteness estimate will be an upper limit only.  See Section \ref{subsubsec:3galdiff} for more details. 

Similarly to the process for the synthetic novae, each of the above 32 Galactic nova light curves was begun at $t_0 = [-20, -19, -18,$ $\cdots, 345]$~days\footnote{Given the overall excellent sampling frequency and extended duration of the Galactic nova light curves, we were able to model outbursts occurring over a full year, starting 20 days before the \hst observing campaign of M51 commenced.} and sampled with the \hst observing cadence.  Linear interpolation was utilized where small gaps in the Galactic nova light curves coincided with one of the M51 \hst epochs.  This provided us with a realistic set of $11,680$ simulated nova light curves.  Then, as we had done for the synthetic novae, we evaluated each of the simulated light curves to determine whether it would have yielded a nova detection using our $V_{606}$-band selection parameters, summarized in Section \ref{sec:data}.

The results of our detection efficiency evaluations for the 32 Galactic novae with the most reliable distances are shown in the left panel of Figure \ref{fig:3deteff}.  
The $M_{peak}$, $t_2$, and mean detection efficiency $\epsilon$ shown for each data point in Figure \ref{fig:3deteff} correspond to the Galactic nova that was used to generate the simulated light curves.  The mean $\epsilon$  
of this sample of 32 Galactic novae, which is devoid of any faint/fast types, {\it is a hard upper limit} to the completeness for our M51 survey: $\epsilon = 69$~percent.

\subsection{Simulation II.  M87 Novae} \label{subsec:m87}

Next to our own Galaxy and M31, the massive elliptical galaxy M87 boasts the largest sample of densely observed nova light curves, the product of a two-month \hst observing campaign \citep{Shara2016}.  Due to the brevity of that survey, approximately half of those light curves are too incomplete for our simulation studies, but we were able to utilize $15$ M87 novae whose observed light curves extend beyond $t_2$.  This dataset provides another independent test of the detection efficiency $\epsilon$ of the M51 survey, though faint/slow M87 novae were almost certainly missed.  As in the case of the Galactic novae, this sample of M87 novae provides an independent {\it upper limit} on our M51 survey's detection efficiency.  

We computed the absolute magnitudes for the M87 novae assuming a distance $d = (16.4 \pm 0.5)$~Mpc, corresponding to a distance modulus of $31.1$~mag \citep{Bird2010}.  
Each of the selected light curves was then begun at $t_0 = [-20, -19, -18, \cdots, 345]$~days and sampled with the \hst observing cadence.  The final yield for the M87 novae was a set of $5,475$ simulated nova light curves, which we then evaluated for detectability within the M51 search parameters.  

Figure \ref{fig:3deteff} (center) shows the results of our M87-based nova detection efficiency evaluations.  With a mean of 34 percent, it falls between that of the Galactic and M31 novae, though much closer to the latter (see below).  
We note that although the M87 survey discovered a number of faint/fast novae \citep{Shara2017ii} which would have significantly reduced the overall mean 
$\epsilon$ of the M51 survey, most of those novae were excluded from our simulations because their light curves either dropped below the \hst detection limit too quickly, or were prematurely cut off during the decline period when the survey ended (the latter also occurred for a few of the brighter novae).  Thus, as noted above, the $\epsilon = 34$~percent 
result for the M87 nova sample, applied to the M51 survey, {\it is an upper limit}.

\subsection{Simulation III.  M31 Novae} \label{subsec:m31}

The most unbiased sample of novae available to us is that of M31.  Unlike our own Milky Way, where dust extinction and reddening severely hampers our ability to detect distant novae in the plane of the Galactic disk, M31 affords us a clearer view of its stellar population.
Given that the distances of these novae are all well determined and internal reddening is minimal, 
it is not surprising that our nearest massive galactic neighbor has yielded a significant number of "faint and fast" novae.  As already noted, faint/fast novae are challenging to detect in external galaxies because they are often too faint to detect via ground-based observations, which are limited by poor weather and seeing, bright moon, and easy-to-manage, "a few nights at a time" block scheduling of telescopes.

For this study, we utilized 29 novae observed during two M31 surveys that are among the most complete and unbiased available, because of their near-daily observing cadences over extended periods of time \citep{Arp1956, Kasliwal2011}.  We also added 30 well-sampled M31 nova light curves observed by the Zwicky Transient Facility (ZTF) between 2019-2021 (\url{https://irsa.ipac.caltech.edu/cgi-bin/Gator/nph-scan?projshort=ZTF}).  We excluded from consideration all novae in those surveys whose outburst peaks may have been missed.  This brought our M31 sample size to 59, our largest extragalactic nova sample by far.  

Absolute magnitudes for each of the M31 novae were calculated using a distance modulus of $24.32$~mag \citep{Wagner-Kaiser2015} and assuming a uniform extinction $E(B-V) = 0.062$ \citep{Schlegel1998}.  Each light curve was initiated at $t_0 = [-20, -19, -18, \cdots, 345]$~days and sampled with the \hst M51 observing cadence.  As with the Galactic novae, interpolation was utilized where small gaps in the light curves coincided with one of the M51 \hst epochs.  We were thus able to simulate a set of $21,535$ realistic light curves using the M31 novae.  Once again, the simulated light curves were evaluated for detectability within the M51 survey.  

\begin{figure*}
\includegraphics[width=1.\textwidth]{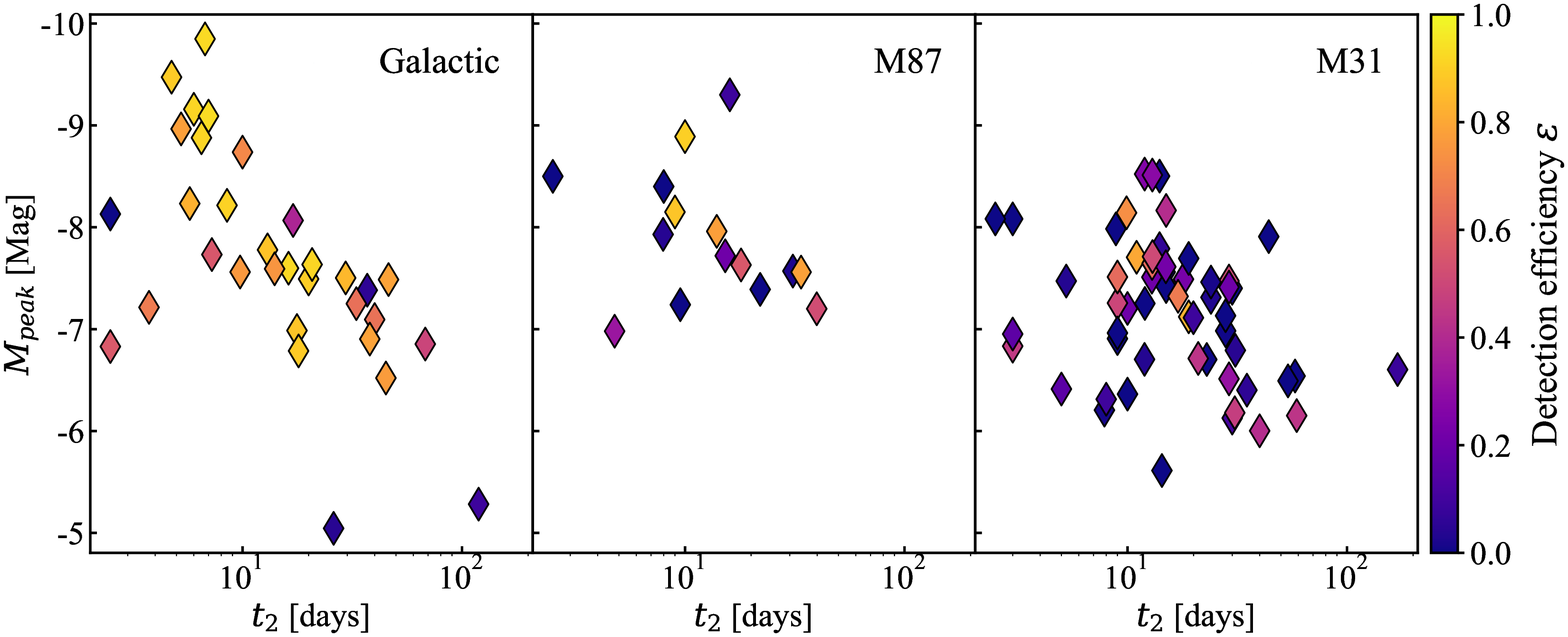}
\caption{Detection efficiency $\epsilon$ for simulated light curves from Galactic (left), M87 (center), and M31 (right) novae.  
See Sections \ref{subsec:gal}--\ref{subsec:m31} in the text for details.
\label{fig:3deteff}}
\end{figure*}

Figure \ref{fig:3deteff} (right panel) shows the detection efficiency $\epsilon$ for the simulated M31 novae; as before, the $M_{peak}$, $t_2$, and 
$\epsilon$ shown in the plot correspond to the "parent" nova that was used to generate the simulated light curves.  Not surprisingly, many of the novae with brighter $M_{peak}$ and longer $t_2$ still had higher overall detection rates than their fainter/faster counterparts.  At 20.5 percent, the mean $\epsilon$ 
for the M31-based simulations was much lower than that of the Galactic-nova-based simulations, as expected given the significant subset of faint/fast novae in the less-biased nova sample of M31.  

\begin{figure}
\includegraphics[width=0.48\textwidth]{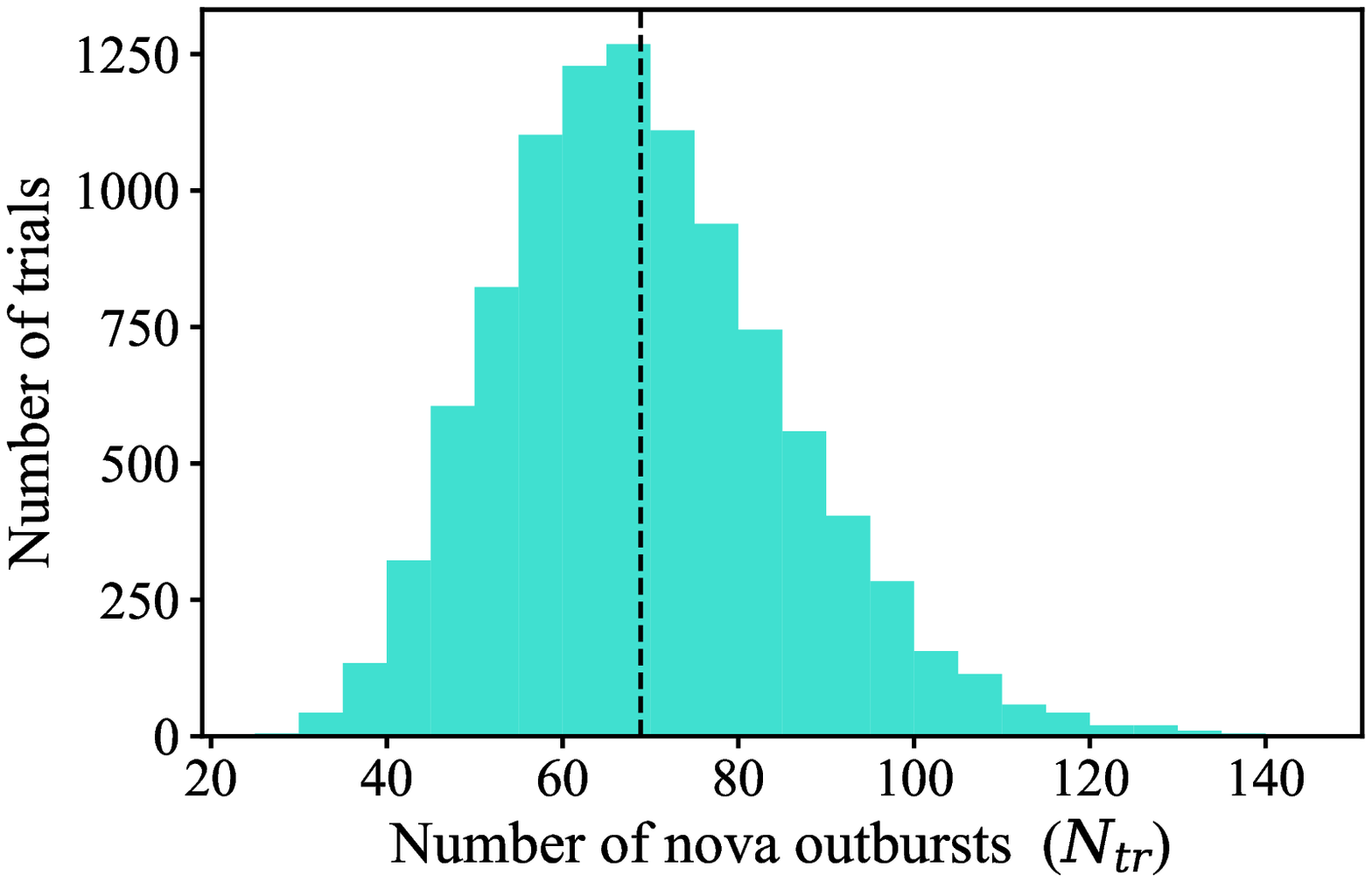}
\caption{Distribution of the "true" number of (randomly sampled) novae, $N_{tr}$, required to recover the number of observed novae, $N_{obs} = 14$, in $10^5$ trials.  The dashed line marks the mean value of $N_{tr} = 69$~novae.  $\frac{N_{obs}}{<N_{tr}>}$ = 20.3 percent, which best approximates the detection efficiency $\epsilon$ of our M51 nova survey if the nova populations of M31 and M51 are not very dissimilar.
\label{fig:Nt}}
\end{figure}

To quantify our completeness we carried out the following set of simulations.  In each trial, we chose one of the M31 nova at random, initiated its outburst on a random day as described above, and analyzed its light curve to determine whether that nova was detectable within the M51 survey, using our $V_{606}$-band selection parameters.  We continued selecting novae at random and evaluating them for detectability until the number detected, $N_{obs}$, reached 14.  The"true" number of novae needed to reach $N_{obs} = 14$ during each trial was recorded as $N_{tr}$.  These trials were repeated $10^5$ times, yielding the distribution of $10^5$ $N_{tr}$ shown in Figure \ref{fig:Nt}.  The mean and 1-sigma ($\pm 34.1$~percent) widths of the $N_{tr}$ distribution are $69^{+19}_{-15}$ novae, corresponding to a nova detection efficiency $\epsilon = 20.3^{+5.6}_{-4.3}$~percent.

The above value is in agreement with the mean detection rate of $\epsilon = 20.5$~percent for the full dataset of M31 simulations.  The mean $\epsilon$ values for the ZTF, \citet{Arp1956}, and \citet{Kasliwal2011} subsets are also consistent, at 20, 22, and 18 percent, respectively.  

\section{Discussion} \label{sec:discussion}

\subsection{Detection Efficiency and the True Nova Rate in M51} \label{subsec:deteff}

\begin{table*}
  \begin{threeparttable}[b]
   \caption
   {Nova detection efficiency from simulations \label{tab:simeff}}
   \centering
   \begin{tabular}{lcccccc}
     \hline\hline 
     Simulation & $<M_{peak}>$\tnote{a} & $<t_{2}>$\tnote{a} & Input & Trials &  \% Detected & Reference\\ 
     & (Abs. mag) & (days) & Novae & & (Mean $\epsilon$) & \\
     \hline
       Synthetic  & -7.5  & 77  & 1071  & 370,566 & $<65^b$ & This work  \\ 
       Galactic  & -7.7  & 22  & 32  & 11,680 & $<69^c$ & 1  \\ 
       M87  & -7.9  & 16  & 15  & 5,475 & $<34^d$ & 2  \\ M31  & -7.1  & 22  & 59  & 21,535 & $20.3^{+5.6}_{-4.3}$ & 3, 4, 5  \\ 
       \hline 
     \end{tabular}
     \begin{tablenotes}
       \item [a] Parameters listed are averaged values from the distributions of observed \textit{input} novae that the sets of simulated light curves were generated from.
       \item [b] This value is a strong upper limit since this simulation assigned equal weight to all simulated novae, including those with simultaneously large $M_{peak}$ and $t_2$; no such novae are known to exist.
       \item [c] This value is a strong upper limit since Galactic surveys are strongly biased against the detection of faint/fast novae (see Section \ref{subsubsec:3galdiff} for more).
       \item [d] This value is an upper limit since the M87 survey was biased against the detection of faint novae, and the fast novae that had been detected were largely excluded from our sample (Section \ref{subsec:m87}).
     \end{tablenotes}
    \textbf{References} (for "input" nova light curves): (1) \cite{Strope2010}; (2) \cite{Shara2016}; (3) \cite{Arp1956}; (4) \cite{Kasliwal2011}; (5) ZTF (\url{https://irsa.ipac.caltech.edu/cgi-bin/Gator/nph-scan?projshort=ZTF}).
  \end{threeparttable}
\end{table*}

The simulations described in section \ref{sec:realsims} indicate that less than a quarter of the novae that underwent an outburst in M51 during the \hst M51 survey were detected, {\it assuming that the nova populations of M51 and M31 are similar.} While it might be tempting to simply lump all of the M31, M87 and Galactic novae of Table 3 into a single simulation, we again emphasize that the most biased sample is that of the Galactic novae, which miss almost all faint, fast novae.  Furthermore, as Figure \ref{fig:lcvar} demonstrates, "fast" novae cannot be quantified by $t_2$ alone, since the light curve decline rate preceding and following the two-magnitude-decline point plays a much larger role in a nova's detectability.  This light curve \textit{shape} effect is evident in Figure \ref{fig:3deteff}, where many of the M31 novae (right panel) have much lower detection efficiency than their Galactic counterparts (left panel) within the same \Mpeak-- $t_2$ parameter space.  That's because the overall \textit{shape} (especially the rapid declines) of many M31 nova light curves 
makes them difficult to detect in our own Galaxy.  Theory and simulations \citep{Yaron2005} predict that a similar population of faint/fast Galactic novae should exist.  These Galactic novae have not yet been discovered because, \textit{at minimum}, months-long, large area, $\sim$~nightly \textit{infrared} (to minimize reddening effects) surveys are required for their detection.  Lumping the M87, M31 and Galactic novae underestimates the number of faint fast novae and overestimates our detection efficiency in M51.  We adopt the detection efficiency $\epsilon$ yielded by our M31 simulations because we consider this estimate to be the most robust, given the relative size and lack of observational bias in that sample.  

\begin{figure}
\includegraphics[width=0.48\textwidth]{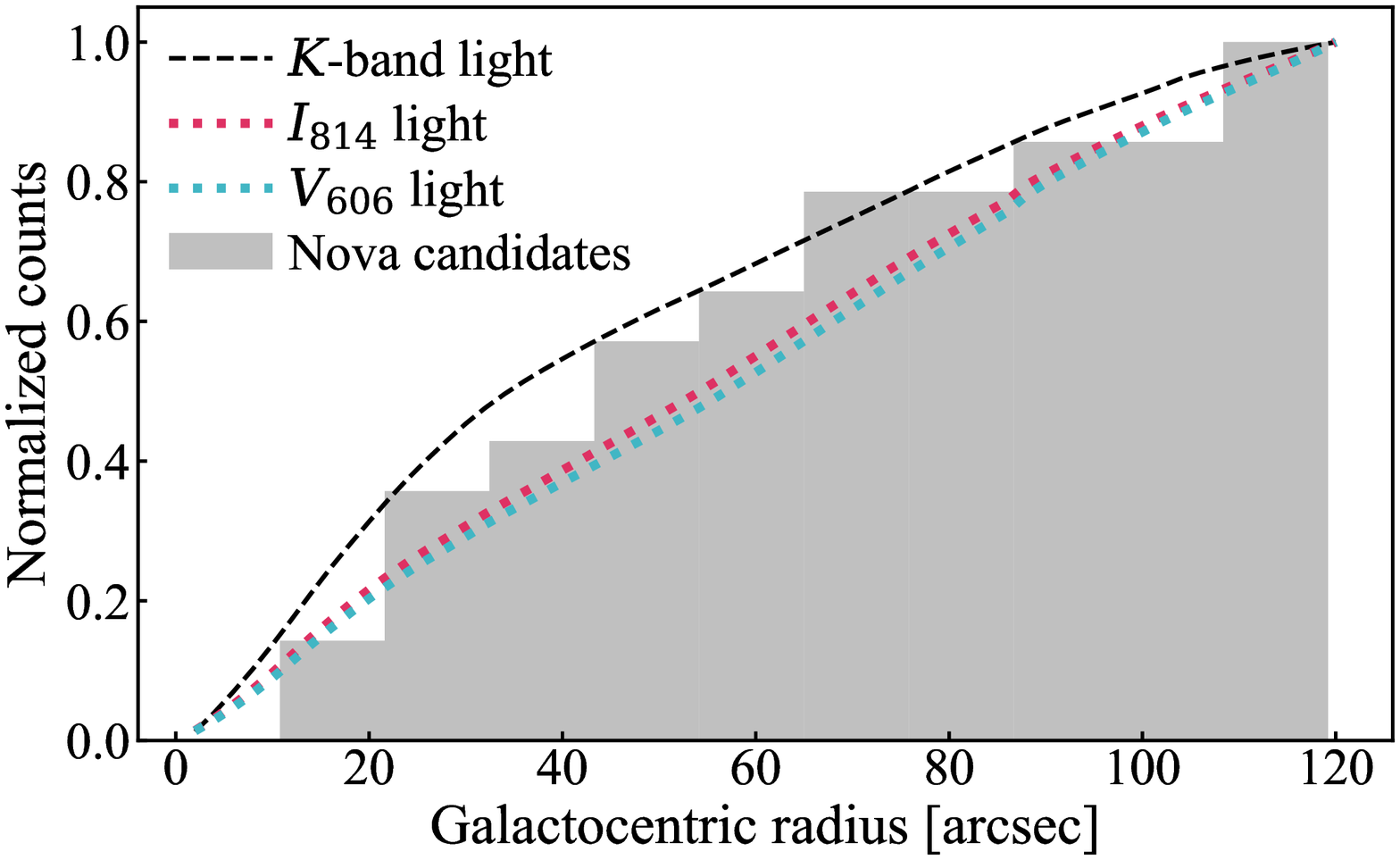}
\caption{Luminosity profile of M51 in \textit{K}-band (black), \textit{I}-band (red), and \textit{V}-band (blue) light.  The cumulative distribution of nova candidates we identified in our survey is shown in grey.  Note that this plot does not show all of M51's light at these radii; it includes the fraction of the galaxy light (and detected novae) that overlap with this survey's \hst-observed area.
\label{fig:raddist}}
\end{figure}

Figure \ref{fig:raddist} shows that the $K$-band light distribution in M51 (which follows the distribution of red giants with masses $\sim$ 1 M$_\odot$) is in good agreement with the spatial distribution of novae in M51 (see Figure \ref{fig:footprint} for the area included in this analysis).  This is consistent with the distribution of novae found in the massive elliptical galaxy M87, which was also shown to follow the $K$-band light \citep{Shara2016}.

As Figure \ref{fig:footprint} shows, our photometric catalog covered a region that encompasses $\sim40$~percent of the $K$-band light from M51.  To calculate this fraction, we first determined the amount of light in the area that consistently remained within the field of view of the \hst survey, a circle of radius $\sim102$~arcsec.  We then subtracted the inner $10$~arcsec, which was excluded from our photometric catalog because its high surface brightness and stellar density presents a challenge for typical photometry tools.  We computed the fraction of M51 light included in our study under the assumption that the total light from M51 is encompassed within a circle of radius $\sim220$~arcsec.  We used a 2MASS K-band image to get the photon counts within the specified areas, yielding a fraction of $\sim40$~percent of the M51 $K$-band light covered by this survey.

We now have in place the three key values needed to derive the M51 nova rate.  These are:

\begin{itemize}
\item the fraction of M51's light consistently observed in the \hst campaign (Figure \ref{fig:footprint}), 40 percent;
\item the determination that the novae "follow the light" in M51 (Figure \ref{fig:raddist}); and
\item the nova detection efficiency for this \hst campaign, $\sim20.3^{+5.6}_{-4.3}$~percent.
\end{itemize}

Then, under the assumption that the underlying CV population in M51 resembles that of  
M31, we conclude that the intrinsic nova rate in M51 is~$172^{+46}_{-37}$~novae yr$^{-1}$.  This M51 nova rate is nearly an order of magnitude higher than previously claimed \citep{Shafter2000}.

\subsection{Nova Rates and Distributions in Different Galaxy Types} \label{subsec:compare}

Although binary population synthesis models have predicted that nova rates ($\nu$, normalized by K-band luminosity $L_K$) should vary with Hubble type \citep{Matteucci2003, Claeys2014, Chen2016}, earlier ground-based surveys found little evidence for this \citep{Shafter2000}.  Given that this survey confirms what several others have already suggested -- that ground-based surveys tend to miss a large fraction of nova outbursts \citep{Shara2016, Mroz2016} -- it is now time to reevaluate the evidence.

Assuming a value of $L_K = 16.6\times10^{10} L_{\odot,K}$ for M51 \citep{Shafter2000} and a rate of $172^{+46}_{-37}$~novae yr$^{-1}$, we find a luminosity-specific nova rate (LSNR) of $\sim10.4^{+2.8}_{-2.2}$~novae yr$^{-1}$/$10^{10} L_{\odot,K}$ for M51.  Both rates are nearly an order of magnitude higher than the ground-based values from \citet{Shafter2000}.  Remarkably, \citet{Shara2016} found a luminosity-specific nova rate of $7.88^{+2.3}_{-2.6}$~novae yr$^{-1}$/$10^{10} L_{\odot,K}$ for M87, within the error range of our \hst-derived rate for M51.  
The similarity of these values is inconsistent with \citet{Chen2016}, who predicted an order-of-magnitude-higher LSNR value for M51-like galaxies than for M87-like elliptical galaxies.  The best evidence now in hand demonstrates that the difference between the LSNRs of M87 and M51 may be much smaller than that predicted.  Note that the M87 nova rate was conservatively calculated, assuming no incompleteness \citep{Shara2016}.  Applying our incompleteness analysis to the M87 \hst survey may well yield an even higher nova rate for that galaxy -- even closer to the LSNR we observe for M51.  

\begin{figure*}
\includegraphics[width=1.\textwidth]{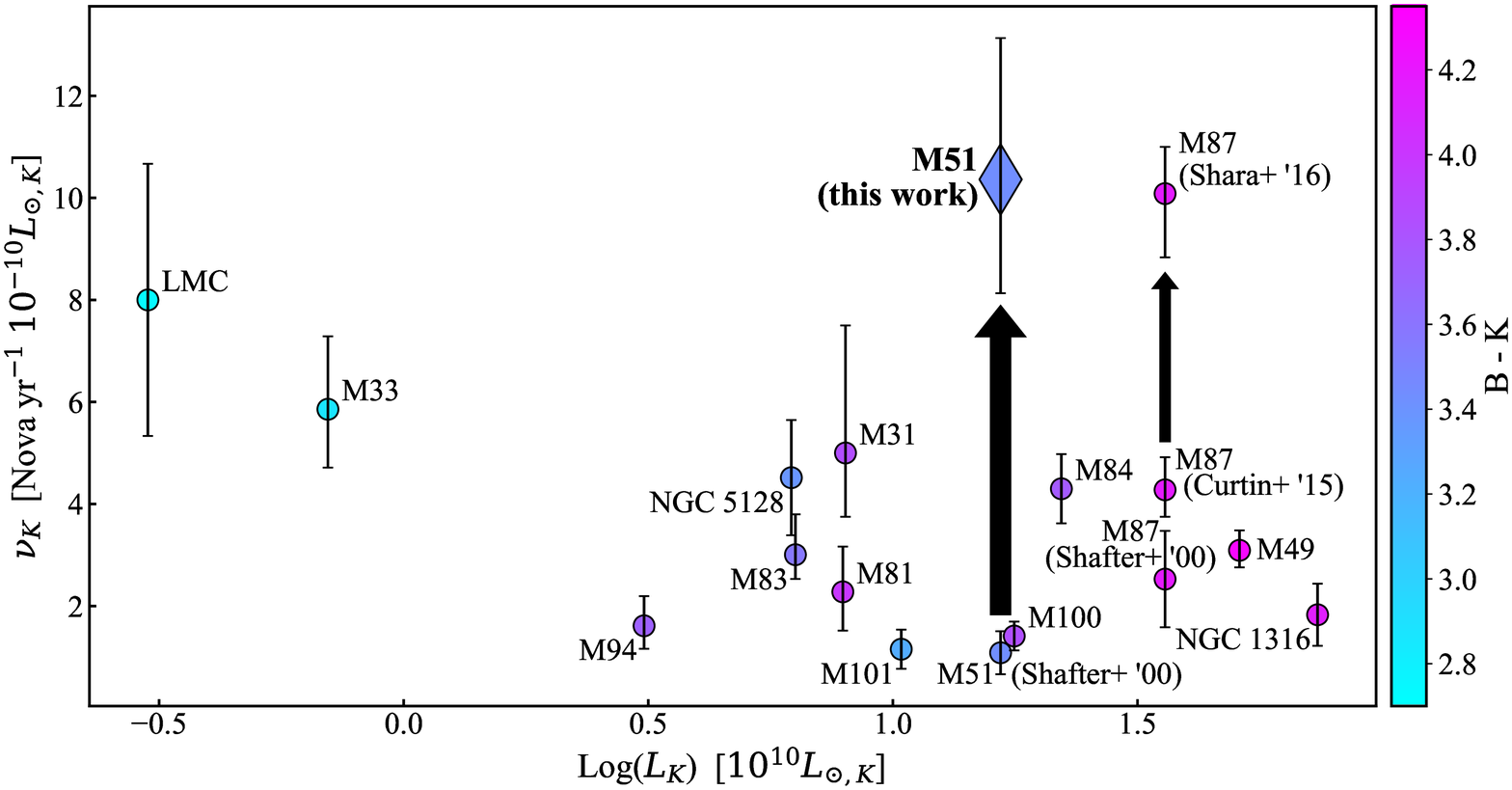}
\caption{Luminosity-specific nova rates for an array of galaxies.  All values are from \citet{Della_Valle2020} (table 7), with the exception of M83 (which was taken from \citet{Shafter2021}), and the two galaxies with \hst-based nova rates: M51 and M87.  For the former, we adopted a \textit{K}-band luminosity of $L_K = 16.6\times10^{10} L_{\odot,K}$ \citep{Shafter2000}.  The larger black arrow highlights the difference between the previously published LSNR and our newly determined rate for M51 (diamond marker).  The smaller arrow denotes the increase in LSNR for M87 established via the high-cadence \hst survey by \citet{Shara2016}, compared to earlier values calculated by \citet{Shafter2000} and \citet{Curtin2015} through ground-based studies.
\label{fig:lsnr}}
\end{figure*}

Figure \ref{fig:lsnr} shows the published luminosity-specific nova rates for 13 different galaxies.  While our newly extrapolated rate for M51 appears to be higher than those of 
the 12 other galaxies, this may not in fact be the case, as none of the other 
12 nova rates 
were derived using our detailed detectability analysis.  Comparable studies of other galaxies may well yield equally high luminosity-specific nova rates by improving incompleteness corrections.  Similarly, high-cadence surveys will also increase nova rates universally, because more frequent observations lower incompleteness.  Such is the case for M87 and the LMC, whose LSNRs are already higher than the 10 remaining galaxies'.  
Unlike most of the rates shown in Figure \ref{fig:lsnr}, the LSNRs for these two galaxies were extrapolated from surveys with near-daily cadence.  Hence they can be considered far more reliable than their sparsely surveyed counterparts -- particularly with respect to faint/fast novae.

\subsection{Additional Uncertainties} \label{subsec:errs}

The following is a discussion of the various uncertainties in our analysis, related to a) the selection of novae, and b) the detection parameters, that were used for the simulations described in Section \ref{sec:realsims}.  

\subsubsection{Nova population differences between galaxies} \label{subsubsec:3galdiff}

Our analysis relies on several assumptions which are not definitive and which impact our final results.  Chief among them is the assumption that the nova population in M51 resembles that of M31.  As noted above, and shown in Table \ref{tab:simeff}, assuming instead a nova population like that in our Galactic sample would result in a significantly higher detection efficacy for this survey, and consequently, a lower derived nova rate estimate.  We repeat that the nova rate determined using the M31 dataset is almost certainly more reliable, given that the M31 sample a) includes faint-fast novae which are conspicuously absent from the Galactic sample, b) is compiled from unbiased\footnote{We sourced the M31 novae from surveys selected for their high cadence and long duration.  Combined with the relatively (compared to our Galaxy) low variation in reddening and accurate distance to M31, which provided us with well-constrained absolute magnitudes, this resulted in the most unbiased sample available, as detailed in Section \ref{subsec:m31}.} surveys, and c) boasts a much larger sample size.   

The survey efficiency results of the simulation based on M87 novae are more consistent with those derived from the M31 dataset than its Galactic counterpart.  However, as Table \ref{tab:simeff} shows, that sample is biased towards the brighter novae.  The larger distance to M87, combined with the relatively short duration of the M87 \hst survey, hampered the detection of well-sampled novae of longer duration and lower $M_{peak}$.  
We also note that as a massive elliptical galaxy with an overall older stellar population, the nova distribution in M87 may be different from that of its spiral counterparts...  which is why we did not simply lump together the M31 and M87 nova samples.

\begin{figure*}
\includegraphics[width=0.8\textwidth]{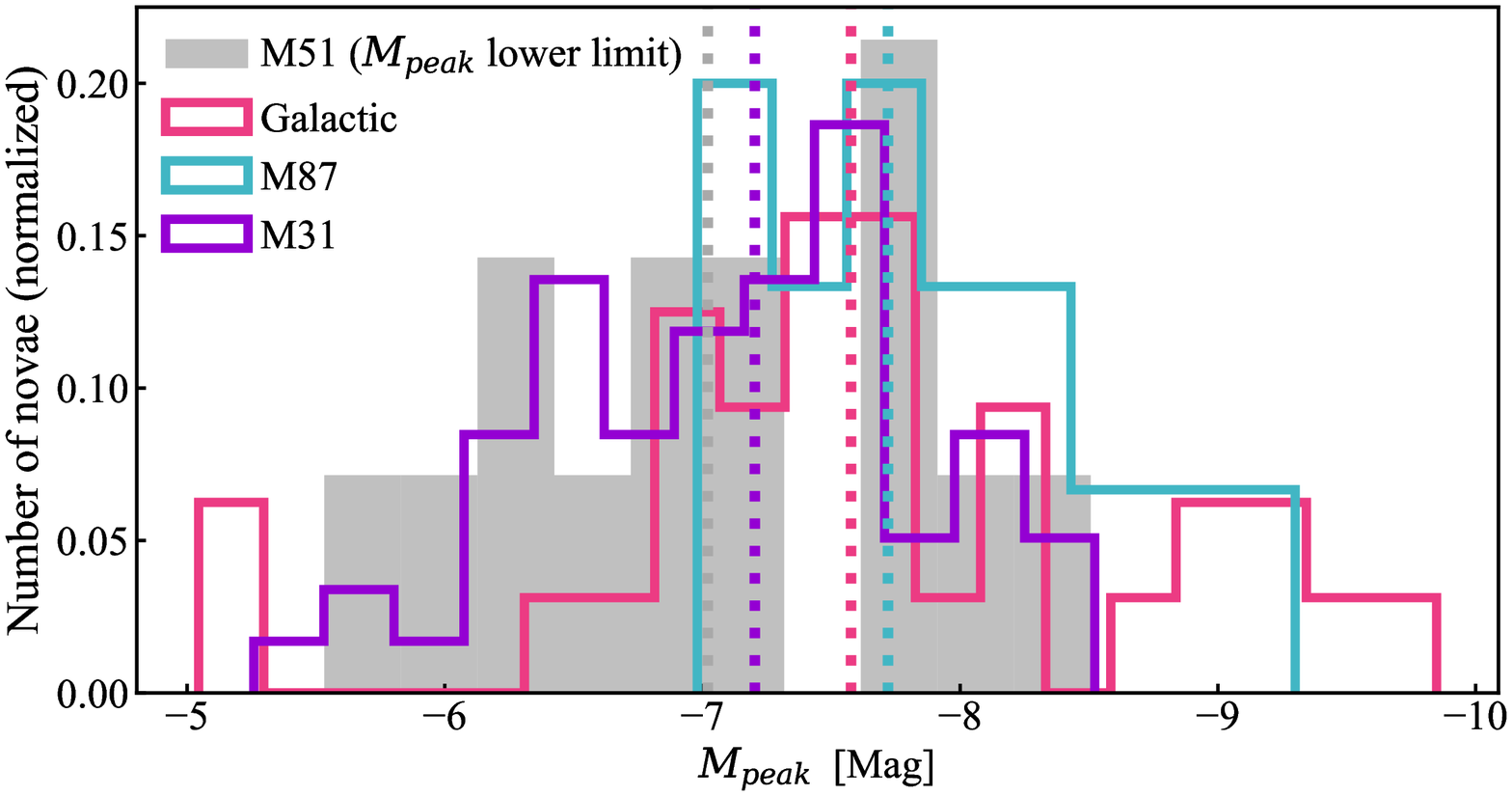}
\caption{\Mpeak distributions for Galactic (pink), M31 (violet), M87 (blue), and M51 (grey) novae.  For M51, all \Mpeak values are lower limits as the outburst "peaks" were likely missed for most novae due to the survey observing cadence.  The dotted lines denote median values.  \label{fig:mpeak_dist}}
\end{figure*}

\begin{figure*}
\includegraphics[width=1.\textwidth]{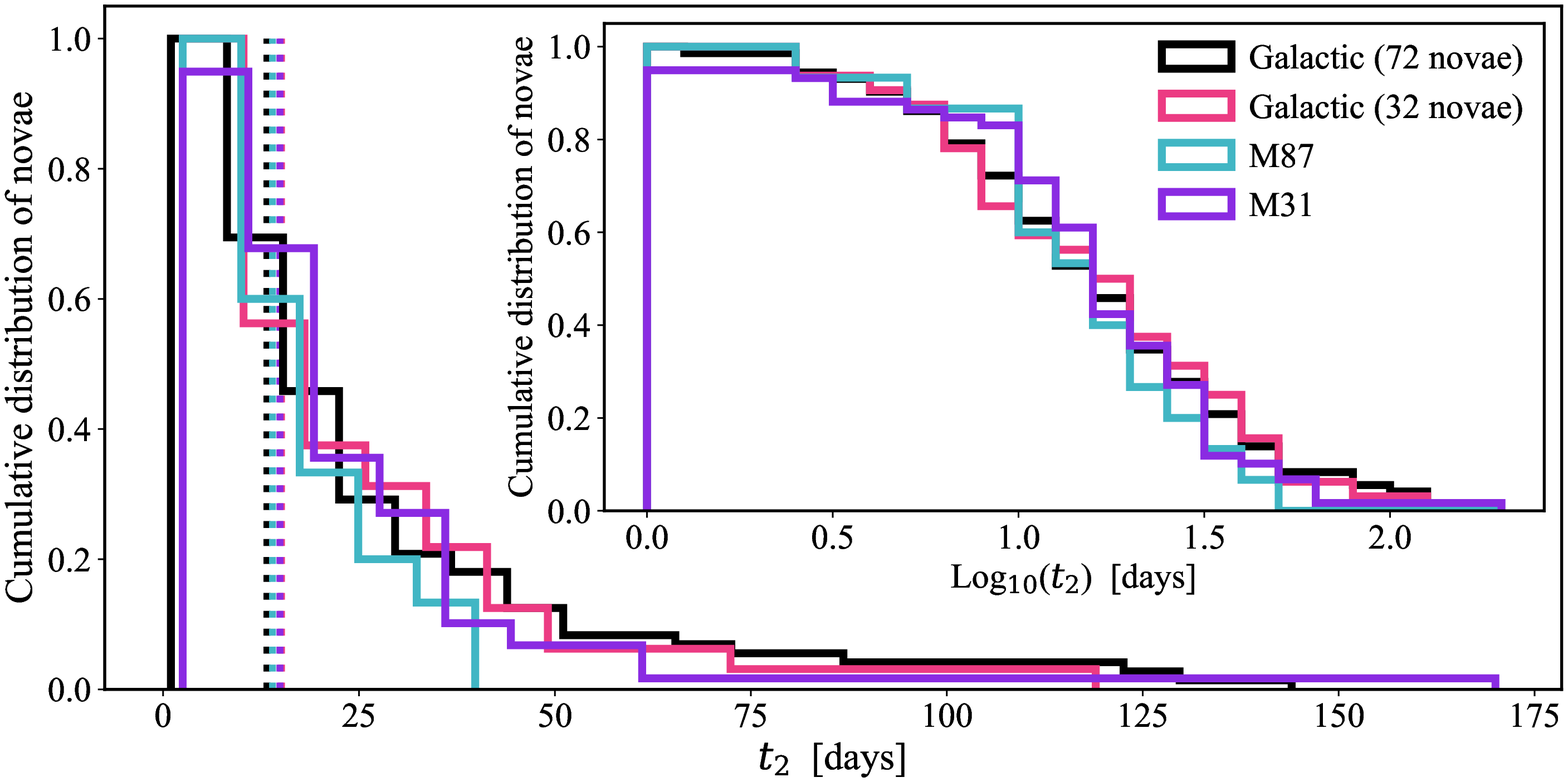}
\caption{Cumulative $t_2$ distributions for Galactic (pink), M31 (violet), and M87 (blue) novae used in the simulations described in Section \ref{sec:realsims}.  A larger sample of Galactic novae, including 40 whose distances are not well constrained, is shown in black.  Dotted lines denote the median values.  The inset shows the distributions in log scale.  
\label{fig:t2_dist}}
\end{figure*}

Figure \ref{fig:mpeak_dist} shows the distributions of \Mpeak for the Milky Way (32 novae), M31 (59 novae), M87 (15 novae), and M51 (14 novae) sample light curves.  Due to the large gaps between the \hst observations of M51, we cannot determine the actual peak luminosities of those novae; we can only determine their lower limits.  
The fainter "tail" of the  
M87 nova distribution is deficient.  We attribute this to observational selection, i.e. the difficulty in detecting such faint novae, even with \hst, at the distance of M87.  In addition, the relatively short duration of that survey prevented the detection of slowly-declining novae, which have lower \Mpeak.  The higher-\Mpeak tail of M51 is missing as well -- again, not surprisingly -- as one might expect from a survey with (on average) 10 day gaps, missing most luminous, fast novae near maximum luminosity.  

Figure \ref{fig:mpeak_dist} also appears to show a dearth of Galactic novae with \Mpeak fainter than $\sim-6.75$.  This is at least partly due to a selection bias that favors the detection of brighter Galactic novae, as described in Section \ref{subsec:gal}.  Note that \citet{Ozdonmez2018} and \citet{Shafter2017} find a mean \Mpeak of $-7.2$ for a subset of Galactic novae.  In the case of the former, novae with well-determined distances were selected; for the latter, it was predominantly novae from the Galactic bulge.  Since our own selection criteria were stricter (requiring well-determined distances and well-sampled light curves), it is not surprising that the mean \Mpeak for our sample is higher (Table \ref{tab:simeff}).  This further supports our claim that the resulting detection efficiency $\epsilon$ for our limited Galactic sample is an overestimate.  In contrast, the mean \Mpeak for our M31 sample is consistent with not only the abovementioned estimates for larger subsets of Galactic novae, but also the entire dataset of observed novae in M31 \citep{Ozdonmez2018, Shafter2017} -- and not just the limited subset of well-sampled novae from unbiased surveys included in our simulation.  This consistency further supports our adaptation of the $\epsilon$ value yielded by the M31 simulations.

Perhaps even more telling are the distributions of $t_2$ in the different galaxies.  Unlike the absolute magnitudes, $t_2$ values are well constrained for 72 Galactic novae, since the latter can be determined from apparent magnitudes, even when the distance to the source is unknown.  As Figure \ref{fig:t2_dist} shows, the overall distribution of $t_2$ for the novae in our M31 dataset is 
similar to that of the Galactic novae.  This is true both for the larger distribution of 72 novae and for the subset of 32 novae (with well-constrained distance estimates) that we utilized in the simulations described in Section \ref{subsec:gal}.  This further indicates that our M31 sample is a good representation of the typical nova population within a barred spiral galaxy.

\subsubsection{Visual identification (or mis-identification) of nova light curves}

Of the approximately 1,000 preliminary nova candidates selected for visual inspection from the full M51 \hst dataset, only 14 light curves were clearly those of transients.  Given the appearance of many of the simulated light curves (described in Section \ref{sec:realsims}), which were based on observed nova samples that included a broader variety of shapes, 
we have good reason to believe that a significant number of novae embedded in the M51 data went unidentified, because, upon review, their light curves did not resemble the canonical nova decline shapes (Section \ref{sec:novae}).  We applied a similar visual test to a subset of the simulated light curves that were marked as "detected" by our nova selection parameters, and found that as many as half of those light curves do not pass the by-eye inspection.  This indicates that our survey detection efficiency could be overestimated by a factor of two -- and consequently, that the intrinsic M51 nova rate could well be double our estimate of $172^{+46}_{-37}$~novae yr$^{-1}$.  Out of an abundance of caution, we decided against incorporating this aspect of our analysis into our results.  But we note that the M51 nova rate of $172^{+46}_{-37}$~novae yr$^{-1}$ is a conservative determination.  

Another factor to consider is that the 14 nova candidates we identified haven't been spectroscopically confirmed.  In the unlikely event that a small subset of them are not actually novae, the extrapolated nova rate for M51 would be reduced.  For example, if two of the 14 sources are \textit{not} in fact novae, the rate would be decreased to $\sim148$~novae yr$^{-1}$ -- which is still far higher than the previously published rate of 18 novae yr$^{-1}$ \citep{Shafter2000}.  However, we consider this scenario highly improbable; of the two classes of variable massive stars that can reach luminosities equivalent to novae, Mira variables are far redder than our nova candidates, and luminous blue variables typically rise in luminosity over months to years, not days, as is the case for our sources.  

\subsubsection{Color limitations}

Among the crucial selection parameters used to winnow nova candidates from the enormous photometric dataset of M51 was the color near maximum light.  Most novae exhibit a "blue" color near maximum light of $V - I < 0.50$~mag, but this property is by no means universal.  Since we lacked sufficient "color" data for most of the novae in our simulation samples, we did not employ this selection criterion for the simulated set.  Therefore, the detection efficiency vis-\`a-vis the actual M51 \hst data may be -- once again -- overestimated, and our derived nova rate consequently \textit{underestimated}.  

\subsubsection{Under- (or over-) representation of faint/fast novae in simulated samples}

The number of observed faint/fast nova light curves available to us is rather small.  
Given the difficulty in detecting such novae, their intrinsic ubiquity is nearly impossible to reliably determine from current data.  In our analysis, we assumed that the novae contained in our M31 dataset constitute a representative sample, although faint/fast novae are likely underrepresented.  We also cannot rule out that such faint/fast novae do not commonly occur in other galaxies, including M51 -- though we consider it far more likely that their detection in M31 and M87 was a result of superior detection efficiency in those surveys (compared to Galactic ones).

\subsubsection{Magnitude uncertainties}

Our analysis assumed that extinction internal to M51 is negligible and approximately uniform.  Similarly, for the M31 novae used in our simulations, we did not account for possible differential extinction.  This could introduce additional errors in magnitude that were not incorporated into our analysis.  However, we found that arbitrarily making all M31 light curves one magnitude fainter had no significant effect on the mean detection efficiency for the M31 simulations.  This indicates that internal extinction effects are likely negligible.

We also note that the novae in our M31 dataset were observed in similar but not identical filters as the M51 \hst novae.  As noted above, varying the luminosity of our sample novae by up to one magnitude had no discernible effect on their overall detectability, so a slight shift in filter wavelength should not significantly influence our results.

\section{Conclusions} \label{sec:conclusion}

We conducted a study of the nova rate in M51 using a year-long \hst survey and realistic simulations to thoroughly test the nova detection efficiency $\epsilon$ of that survey.  Our incompleteness simulations modelled the detectability of well-observed M31 novae with unprecedented detail.  This allowed us to extrapolate M51's intrinsic nova rate with an unparalleled degree of accuracy -- under the assumption that the nova population in M51 resembles that of M31.  We find that the nova rate in M51 is $\approx172^{+46}_{-37}$~novae yr$^{-1}$, and its luminosity-specific nova rate is $\sim10.4^{+2.8}_{-2.2}$~novae yr$^{-1}/10^{10} L_{\odot,K}$.  Both these rates are nearly an order of magnitude higher than the previous published value based on ground-based observations \citep{Shafter2000}.  In contradiction to theoretical predictions of order-of-magnitude differences in LSNRs between elliptical and spiral galaxies, 
M51 and M87 appear to display $\sim$ comparable LSNRs.  The novae in M51 closely follow the $K$-band light distribution in that galaxy, similar to the novae of M87.

\section*{Acknowledgements}

This work was supported by the National Science Foundation Graduate Research Fellowship Program under Grant No. DGE 2036197.  Any opinions, findings, and conclusions or recommendations expressed in this material are those of the authors and do not necessarily reflect the views of the National Science Foundation.  CC and PvD acknowledge support from NASA Grant HST-GO-14704.001.  We would like to thank Jay Strader for his valuable critiques and suggestions, which helped to significantly improve an earlier draft of this paper.

\vspace{5mm}
\textit{facilities:}  HST(ACS), AAVSO, ZTF 
\section*{Data Availability}

The M51 photometry and simulation data underlying this article will be shared on reasonable request to the first author.  The other photometry data are publicly available from the references provided in Table \ref{tab:simeff}.


\bibliographystyle{mnras}
\bibliography{M51} 


\clearpage
\onecolumn

\appendix

\section{Novae used in simulations} \label{sec:novlists}

Some of the Galactic novae listed in this table have two or more values of \Mpeak and $t_2$ reported in the literature.  Where available, we took \Mpeak from the AAVSO data compiled by \citet{Strope2010}, provided that the sampling was $\sim$~daily around the clearly observed peak.  We define $t_2$ as the time it took for the nova to decline from \Mpeak by two magnitudes.  Given that the nova luminosity can change by as much as $1-2$~mag over a timescale of hours, it is not surprising that near-simultaneous observations yielded different values.  For consistency, and to reflect the fact that the \hst M51 epochs were $2.2$~ksec each (and therefore could have easily missed the absolute outburst "peak" even if the observation occurred on the day of peak brightness), we utilized the AAVSO values even when a higher \Mpeak was reported elsewhere.  

\vspace{0.5cm}

\setlength{\tabcolsep}{30pt}
\begin{longtable}[c]{c c c c}
 \caption{Novae Used in Simulations.\label{simnovae}} \\
 \hline\hline
 Nova & $m_{peak}$ & $t_2$ & Filter\\ 
  & (mag) & (days) \\ [1ex] 
 \hline
 \endfirsthead
 Nova & $m_{peak}$ & $t_2$ & Filter \\ 
 \hline
 \endhead
 \hline
 \endfoot

 \hline
 \multicolumn{4}{c}{M31 Novae} \\
 \hline
M31 (Arp '56) 1  & 16.43  & 3.0  &  $m_{pg}$  \\ 
M31 (Arp '56) 2  & 16.43  & 2.5  &  $m_{pg}$  \\ 
M31 (Arp '56) 4  & 18.15  & 10.0  &  $m_{pg}$  \\ 
M31 (Arp '56) 5  & 15.99  & 12.0  &  $m_{pg}$  \\ 
M31 (Arp '56) 6  & 16.37  & 9.9  &  $m_{pg}$  \\ 
M31 (Arp '56) 7  & 16.01  & 14.0  &  $m_{pg}$  \\ 
M31 (Arp '56) 8  & 17.0  & 9.0  &  $m_{pg}$  \\ 
M31 (Arp '56) 10  & 17.04  & 5.3  &  $m_{pg}$  \\ 
M31 (Arp '56) 12  & 16.0  & 13.0  &  $m_{pg}$  \\ 
M31 (Arp '56) 13  & 17.04  & 29.0  &  $m_{pg}$  \\ 
M31 (Arp '56) 15  & 16.9  & 15.0  &  $m_{pg}$  \\ 
M31 (Arp '56) 16  & 17.26  & 12.0  &  $m_{pg}$  \\ 
M31 (Arp '56) 17  & 17.2  & 24.0  &  $m_{pg}$  \\ 
M31 (Arp '56) 19  & 17.72  & 31.0  &  $m_{pg}$  \\ 
M31 (Arp '56) 20  & 17.19  & 17.0  &  $m_{pg}$  \\ 
M31 (Arp '56) 21  & 17.4  & 20.0  &  $m_{pg}$  \\ 
M31 (Arp '56) 23  & 17.38  & 28.0  &  $m_{pg}$  \\ 
M31 (Arp '56) 24  & 17.81  & 23.0  &  $m_{pg}$  \\ 
M31 (Arp '56) 25  & 17.56  & 3.0  &  $m_{pg}$  \\ 
M31 (Arp '56) 26  & 18.0  & 29.0  &  $m_{pg}$  \\ 
M31 (Arp '56) 30  & 18.02  & 53.9  &  $m_{pg}$  \\  
 \hline
M31N 2007-10a  & 17.55  & 9.0  &  \textit{g}  \\
M31N 2008-07b  & 18.9  & 14.4  &  \textit{g}  \\ 
M31N 2008-08c  & 17.1  & 29.0  &  \textit{g}  \\
M31N 2008-09a  & 17.8  & 21.0  &  \textit{g}  \\
M31N 2008-09c  & 16.8  & 13.0  &  \textit{g}  \\
M31N 2008-10b  & 18.1  & 5.0  &  \textit{g}  \\ 
M31N 2008-11a  & 18.2  & 8.0  &  \textit{g}  \\ 
M31N 2008-12b  & 17.05  & 24.0  &  \textit{g}  \\  
 \hline
ZTF21aagkzve  & 17.26  & 9.0  &  ZTF-\textit{g}  \\ 
ZTF21aagkzve  & 17.09  & 15.0  &  ZTF-\textit{r}  \\ 
ZTF21acbcfmh  & 16.53  & 8.9  &  ZTF-\textit{g}  \\ 
ZTF21acbcfmh  & 16.35  & 15.0  &  ZTF-\textit{r}  \\ 
ZTF21abjiotr  & 19.25  & --  &  ZTF-\textit{g}  \\ 
ZTF21abjiotr  & 18.39  & 30.0  &  ZTF-\textit{r}  \\
ZTF20acstbfh  & 17.68  & 3.0  &  ZTF-\textit{g}  \\ 
ZTF20acplkub  & 18.89  & --  &  ZTF-\textit{g}  \\ 
ZTF20acplkub  & 18.31  & 7.9  &  ZTF-\textit{r}  \\
ZTF20acoqrpm  & 16.81  & 11.0  &  ZTF-\textit{g}  \\
ZTF20acgigfo  & 17.6  & 9.0  &  ZTF-\textit{g}  \\ 
ZTF20acgigfo  & 17.0  & 12.9  &  ZTF-\textit{r}  \\ 
ZTF20acfucwr  & 16.95  & 14.9  &  ZTF-\textit{g}  \\ 
ZTF20acfucwr  & 17.02  & 18.0  &  ZTF-\textit{r}  \\ 
ZTF20abqhsxb  & 17.81  & 12.0  &  ZTF-\textit{r}  \\ 
ZTF19acxrihd  & 18.11  & 35.0  &  ZTF-\textit{g}  \\ 
ZTF19acxrihd  & 17.97  & 58.0  &  ZTF-\textit{r}  \\ 
ZTF19adakuos  & 17.53  & 28.0  &  ZTF-\textit{g}  \\ 
ZTF19adakuos  & 17.39  & 19.0  &  ZTF-\textit{r}  \\
ZTF19acqprad  & 18.51  & 40.0  &  ZTF-\textit{g}  \\ 
ZTF19acqprad  & 18.36  & 59.0  &  ZTF-\textit{r}  \\ 
ZTF19acnfsij  & 18.33  & 30.8  &  ZTF-\textit{r}  \\ 
ZTF19acgfhfd  & 17.11  & 30.0  &  ZTF-\textit{g}  \\ 
ZTF19acgfhfd  & 16.82  & 19.0  &  ZTF-\textit{r}  \\ 
ZTF19acfsteg  & 16.72  & 14.0  &  ZTF-\textit{g}  \\ 
ZTF19acfsteg  & 16.6  & 44.0  &  ZTF-\textit{r}  \\ 
ZTF19abirmkt  & 18.12  & --  &  ZTF-\textit{g}  \\ 
ZTF19abirmkt  & 17.91  & 170.0  &  ZTF-\textit{r}  \\ 
ZTF19abfvpjh  & 17.31  & 10.0  &  ZTF-\textit{g}  \\ 
ZTF19abfvpjh  & 16.87  & 12.9  &  ZTF-\textit{r}  \\  [1.ex]
 \hline 
 \multicolumn{4}{c}{M87 Novae} \\
 \hline
1  & 21.84  & 16.0 &  \textit{V} \\ 
2  & 22.25  & 10.0 &  \textit{V} \\ 
3  & 22.64  & 2.5 &  \textit{V} \\ 
4  & 22.74  & 8.0 &  \textit{V} \\ 
5  & 22.99  & 9.0 &  \textit{V} \\ 
6  & 23.18  & 14.0 &  \textit{V} \\ 
7  & 23.21  & 8.0 &  \textit{V} \\ 
9  & 23.42  & 15.2 &  \textit{V} \\ 
10  & 23.51  & 18.0 &  \textit{V} \\ 
11  & 23.57  & 31.0 &  \textit{V} \\ 
12  & 23.58  & 33.8 &  \textit{V} \\ 
15  & 23.75  & 22.0 &  \textit{V} \\ 
19  & 23.9  & 9.5 &  \textit{V} \\ 
20  & 23.94  & 39.8 &  \textit{V} \\ 
23  & 24.16  & 4.8 &  \textit{V} \\[1.ex]
 \hline
  \multicolumn{4}{c}{Galactic Novae} \\
 \hline
V356 Aql  & 6.97  & 37.0  &  \textit{V}  \\ 
V603 Aql  & -0.5  & 5.8  &  \textit{V}  \\ 
V1370 Aql  & 7.7  & 9.8  &  \textit{V}  \\ 
V1494 Aql  & 4.11  & 8.5  &  \textit{V}  \\ 
V705 Cas  & 5.69  & 33.0  &  \textit{V}  \\ 
V723 Cas  & 7.08  & 17.0  &  \textit{V}  \\ 
V842 Cen  & 4.9  & 29.5  &  \textit{V}  \\ 
V476 Cyg  & 1.94  & 7.2  &  \textit{V}  \\ 
V1330 Cyg  & 9.9  & 119.0  &  \textit{V}  \\ 
V1974 Cyg  & 4.27  & 16.2  &  \textit{V}  \\ 
HR Del  & 3.58  & 68.0  &  \textit{V}  \\ 
DN Gem  & 3.58  & 13.0  &  \textit{V}  \\ 
DQ Her  & 1.56  & 40.0  &  \textit{V}  \\ 
V446 Her  & 4.83  & 17.7  &  \textit{V}  \\ 
V533 Her  & 3.0  & 20.0  &  \textit{V}  \\ 
CP Lac  & 2.02  & 6.0  &  \textit{V}  \\ 
DK Lac  & 5.9  & 18.0  &  \textit{V}  \\ 
IM Nor  & 7.84  & 26.0  &  \textit{V}  \\ 
RS Oph  & 4.94  & 6.8  &  \textit{V}  \\ 
GK Per  & 0.19  & 7.0  &  \textit{V}  \\ 
RR Pic  & 0.95  & 14.0  &  \textit{V}  \\ 
CP Pup  & 0.7  & 4.8  &  \textit{V}  \\ 
T Pyx  & 6.77  & 45.0  &  \textit{V}  \\ 
V3890 Sgr  & 8.05  & 5.2  &  \textit{V}  \\ 
U Sco  & 7.7  & 2.5  &  \textit{V}  \\ 
V992 Sco  & 7.7  & 10.0  &  \textit{V}  \\ 
FH Ser  & 4.5  & 46.2  &  \textit{V}  \\ 
V382 Vel  & 2.77  & 6.5  &  \textit{V}  \\ 
NQ Vul  & 6.19  & 2.5  &  \textit{V}  \\ 
PW Vul  & 6.41  & 3.8  &  \textit{V}  \\ 
QU Vul  & 5.33  & 20.8  &  \textit{V}  \\ 
QV Vul  & 7.13  & 38.0  &  \textit{V}  \\ [0.5ex]
\end{longtable}

\vspace{15cm}

\section{Novae in Difference Images ("Postage Stamps")} \label{sec:novim}

\begin{sidewaysfigure*}
\begin{center}
\centering
  \subfloat[Nova 13]{\includegraphics[width=0.49\textheight]{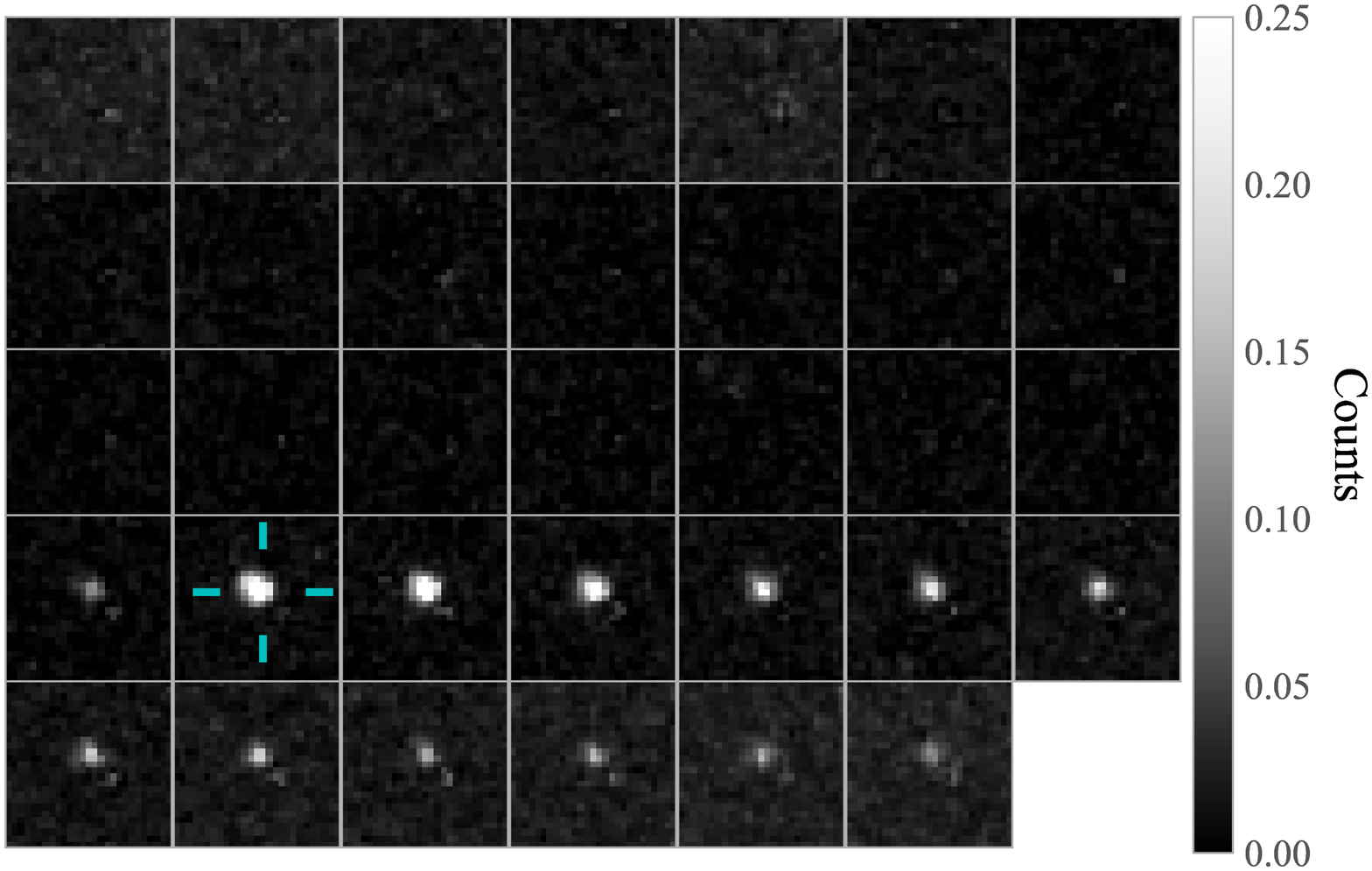}\label{fig:m}} 
  \hfill
  \subfloat[Nova 14]{\includegraphics[width=0.49\textheight]{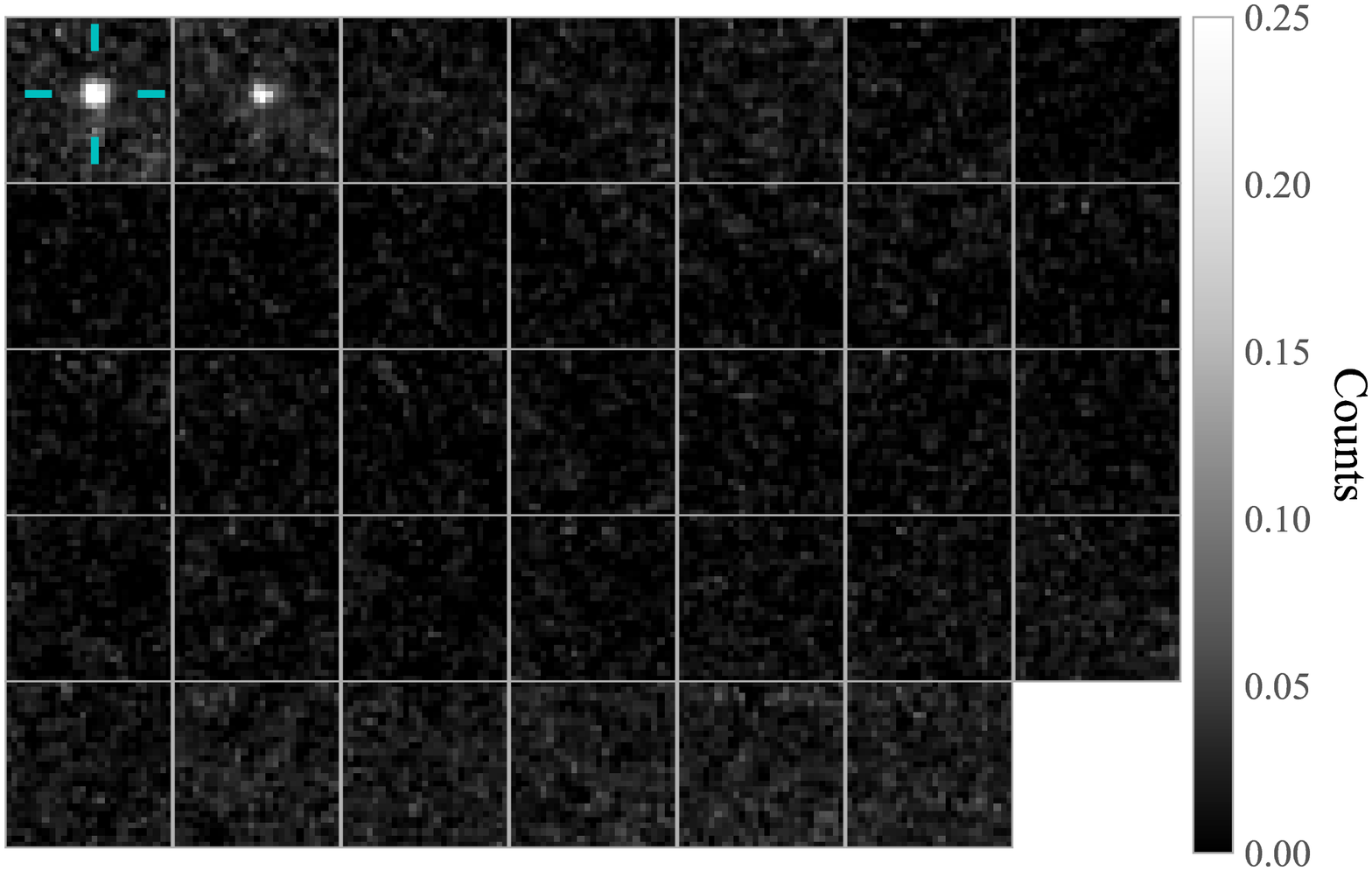}\label{fig:n}}
\caption{$1\arcsec\times1\arcsec$ $V_{606}$ difference image cutouts of the 14 nova candidates in M51 for the 34 epochs of the \hst observing campaign.  Images corresponding to the epochs of observed peak luminosity are marked with cyan ticks.\label{fig:psdiff}}
\end{center}
\end{sidewaysfigure*}

Figure \ref{fig:psdiff} shows the 14 novae that we identified in M51, as they appeared in each observing epoch.  To create these images, we first aligned the four dithered ACS images from each epoch using the \textsc{TweakReg}  
WCS alignment task and drizzled them together with \textsc{DrizzlePac}, 
following the process described in \citet{Hoffmann2021}.  Separately, we created a composite image of M51 by "drizzling" together the individual images from all 34 epochs, again following the procedure described in \citet{Hoffmann2021}.  We then subtracted the composite image from each of the 34 individual drizzled images, creating the difference images shown below.  This process allows us to clearly observe the appearance and decline of the novae.  Each "postage stamps" cutout is 1 arcsec x 1 arcsec 
in size.  
Images corresponding to the epochs of observed peak luminosity are marked with cyan ticks.

\captionsetup[subfigure]{labelformat=empty}
\begin{sidewaysfigure*}
\begin{center}
\centering
  \subfloat[Nova 1]{\includegraphics[width=0.49\textheight]{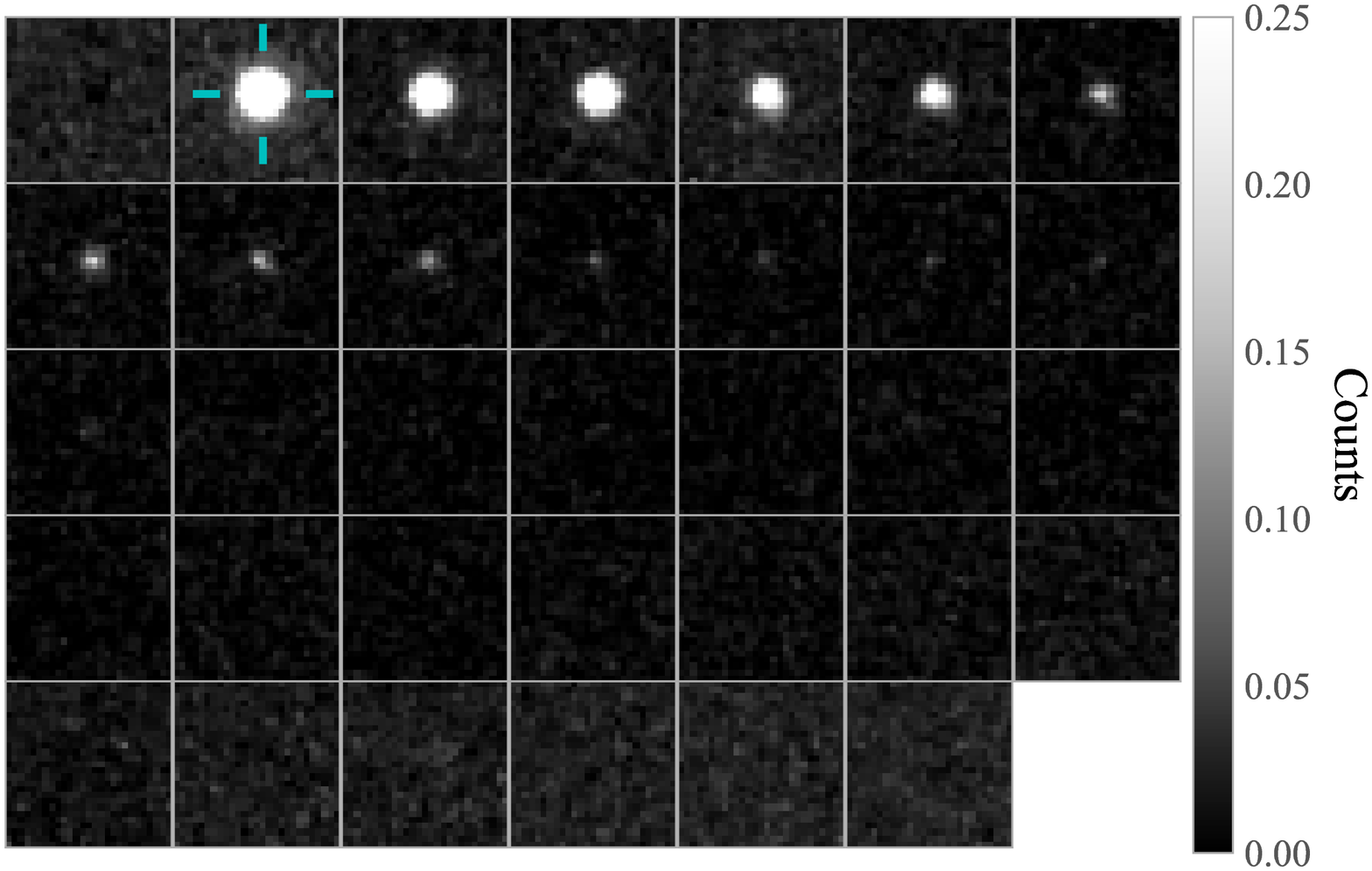}\label{fig:a}} 
  \hfill
  \subfloat[Nova 2]{\includegraphics[width=0.49\textheight]{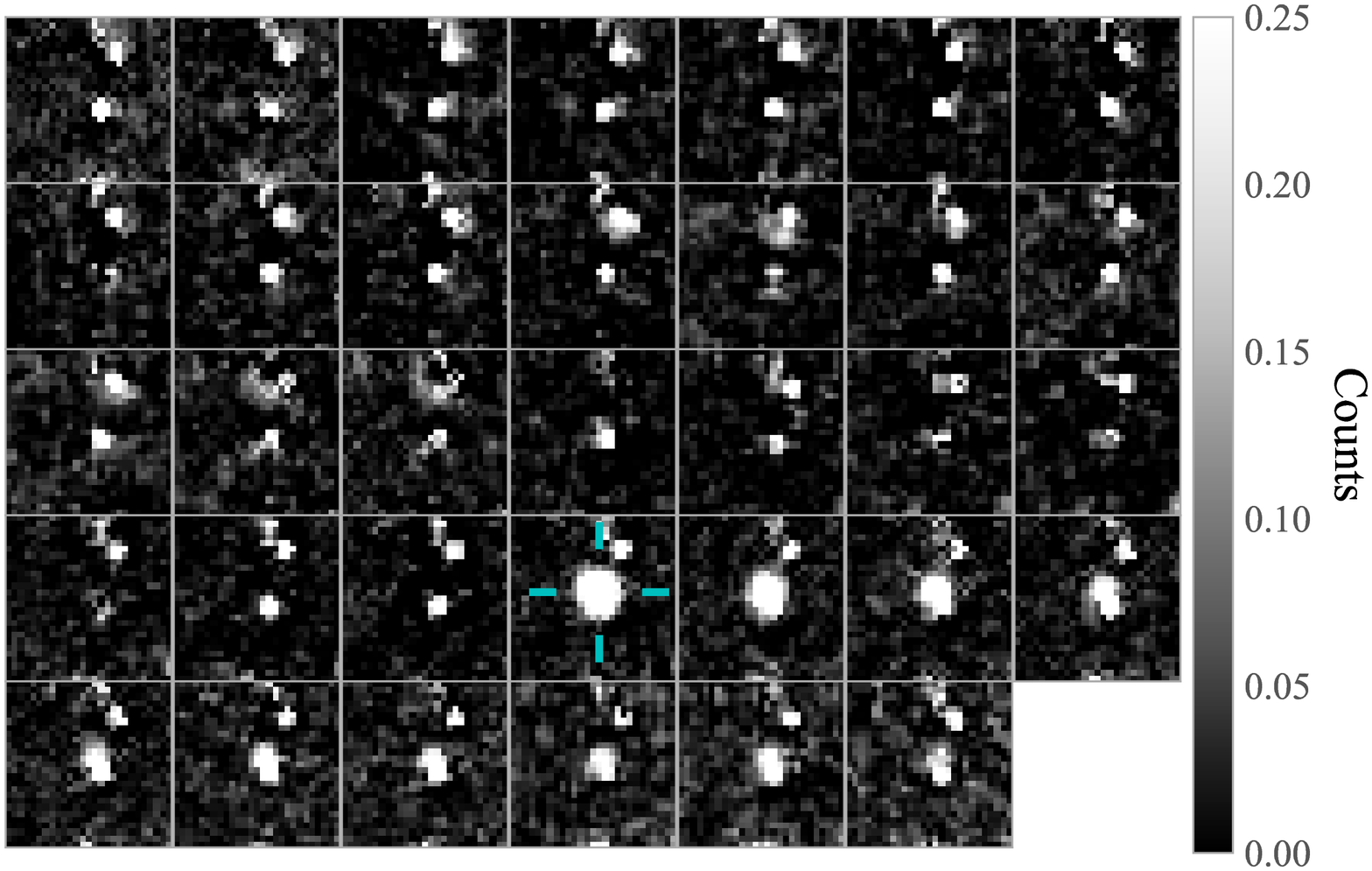}\label{fig:b}}
  \vfill
  \subfloat[Nova 3]{\includegraphics[width=0.49\textheight]{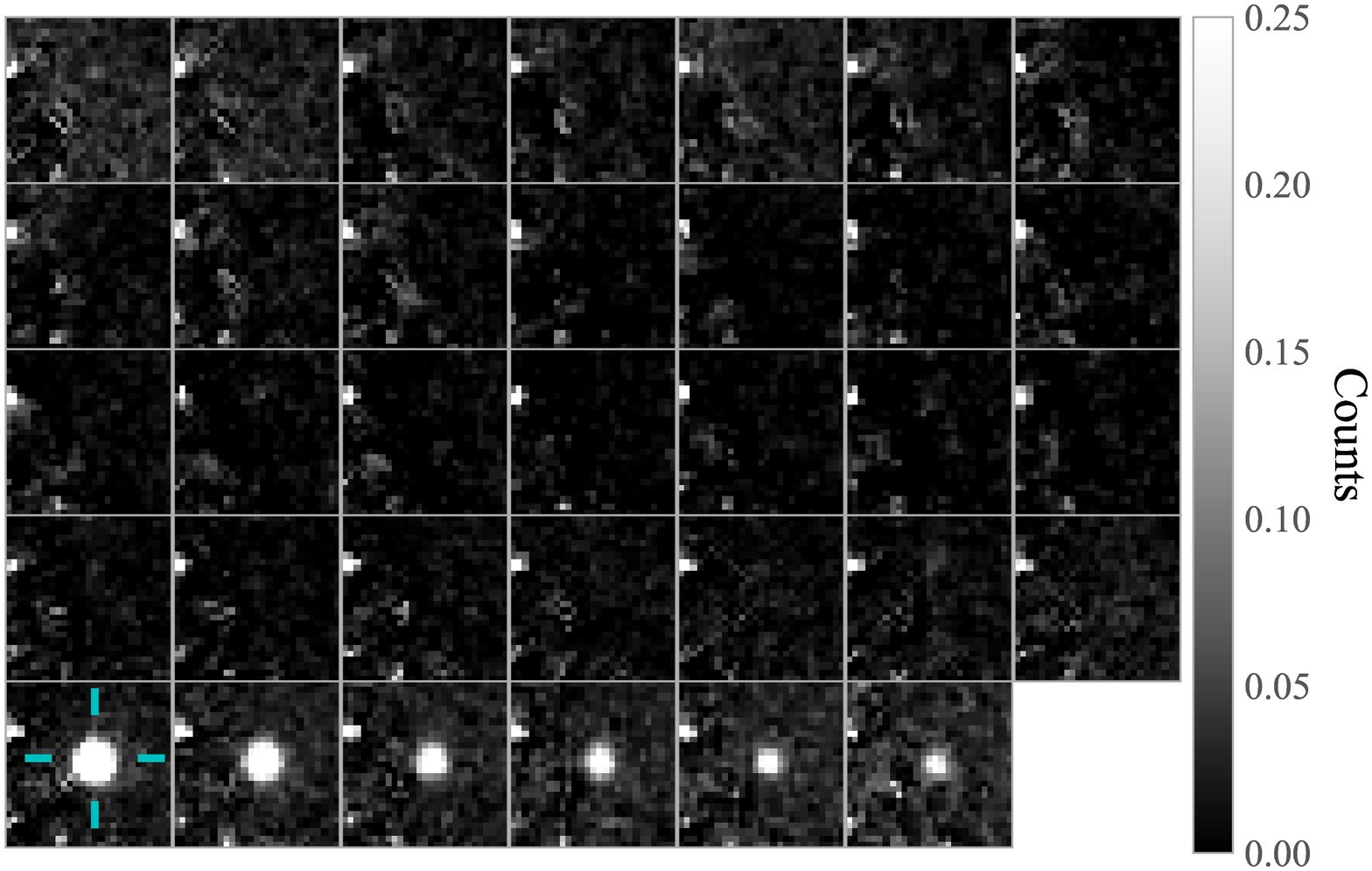}\label{fig:c}} 
  \hfill
  \subfloat[Nova 4]{\includegraphics[width=0.49\textheight]{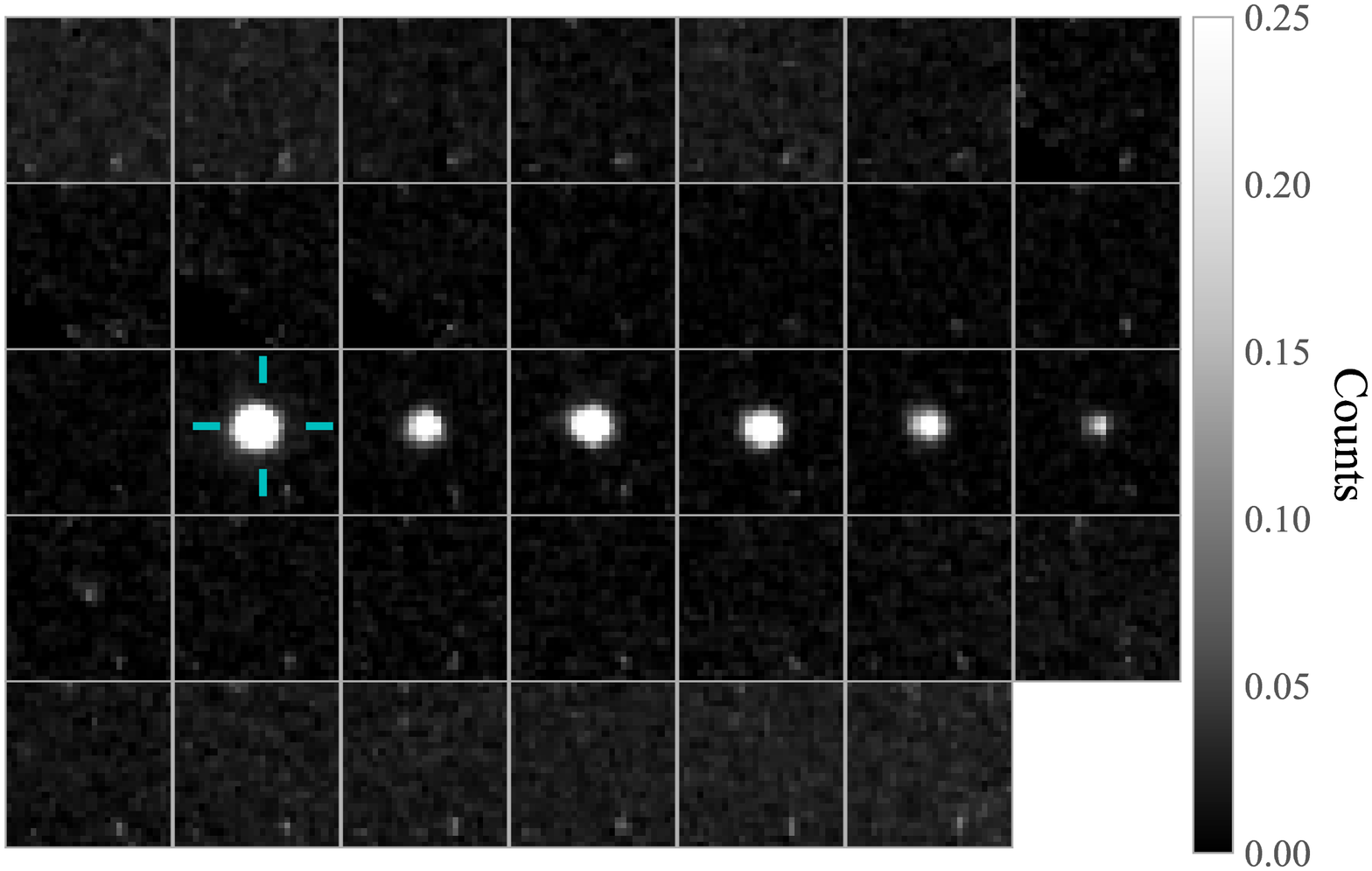}\label{fig:d}}
\end{center}
\end{sidewaysfigure*}

\begin{sidewaysfigure*}
\begin{center}
\centering
  \subfloat[Nova 5]{\includegraphics[width=0.49\textheight]{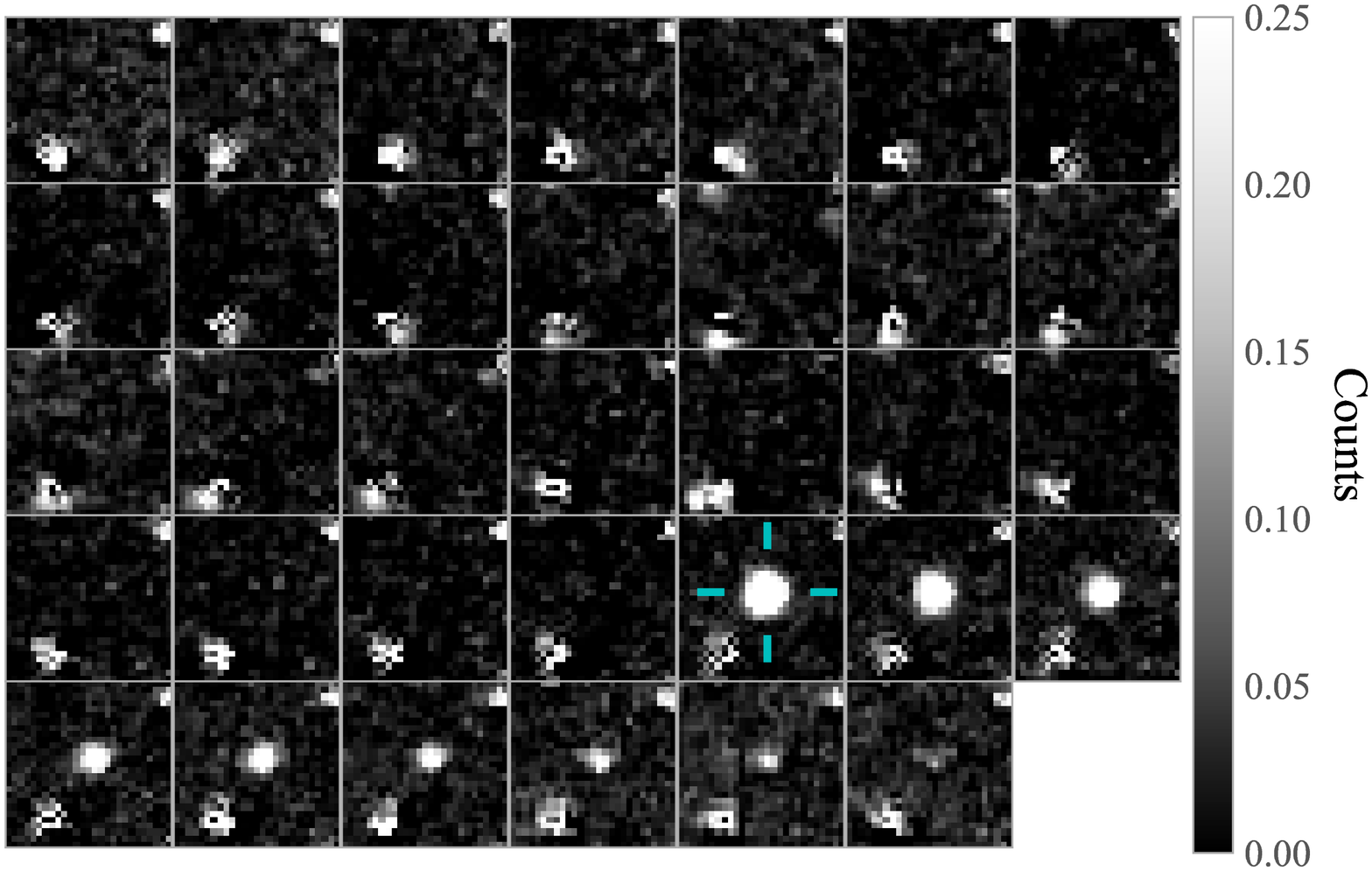}\label{fig:e}} 
  \hfill
  \subfloat[Nova 6]{\includegraphics[width=0.49\textheight]{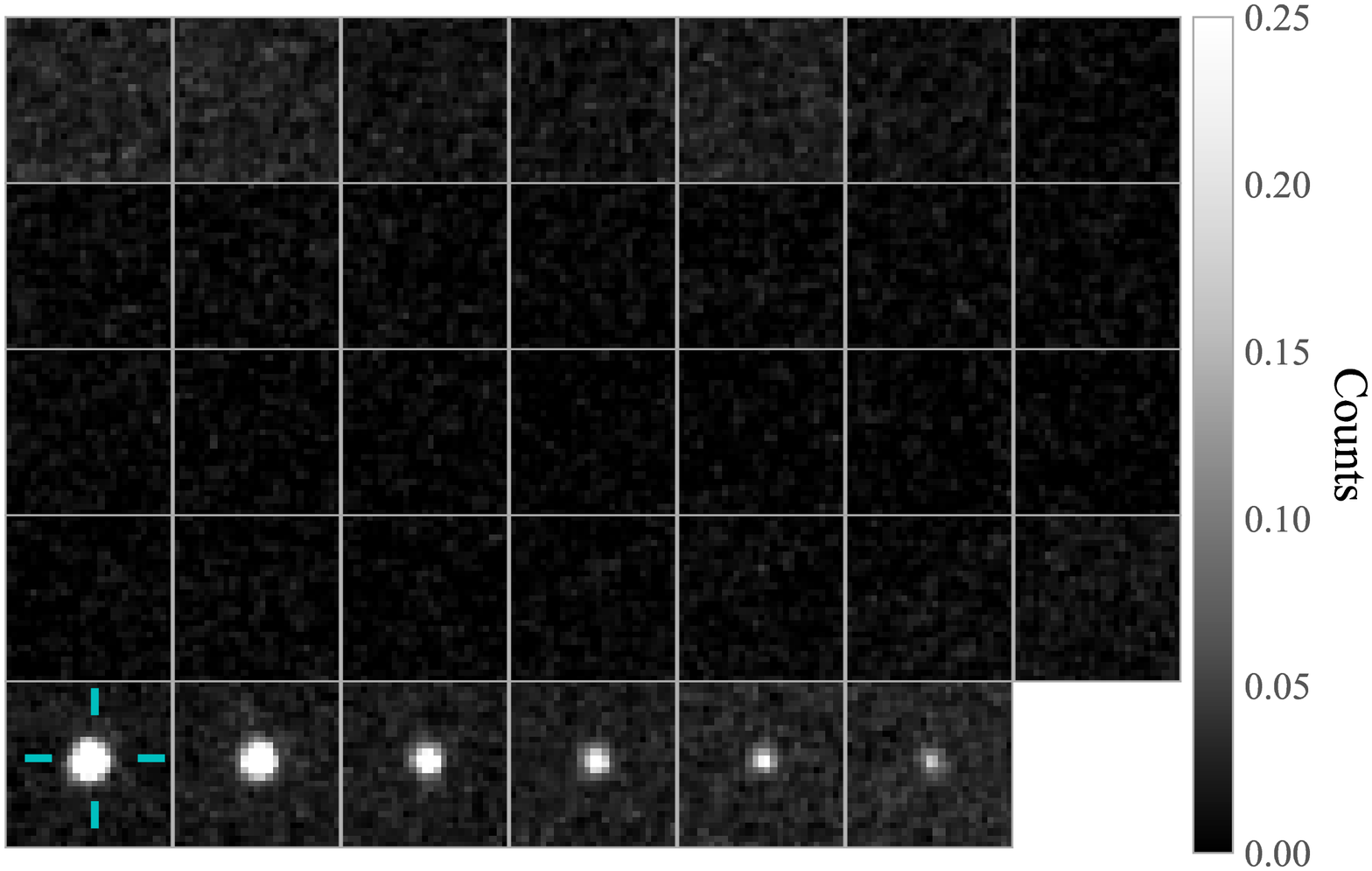}\label{fig:f}}
  \vfill
  \subfloat[Nova 7]{\includegraphics[width=0.49\textheight]{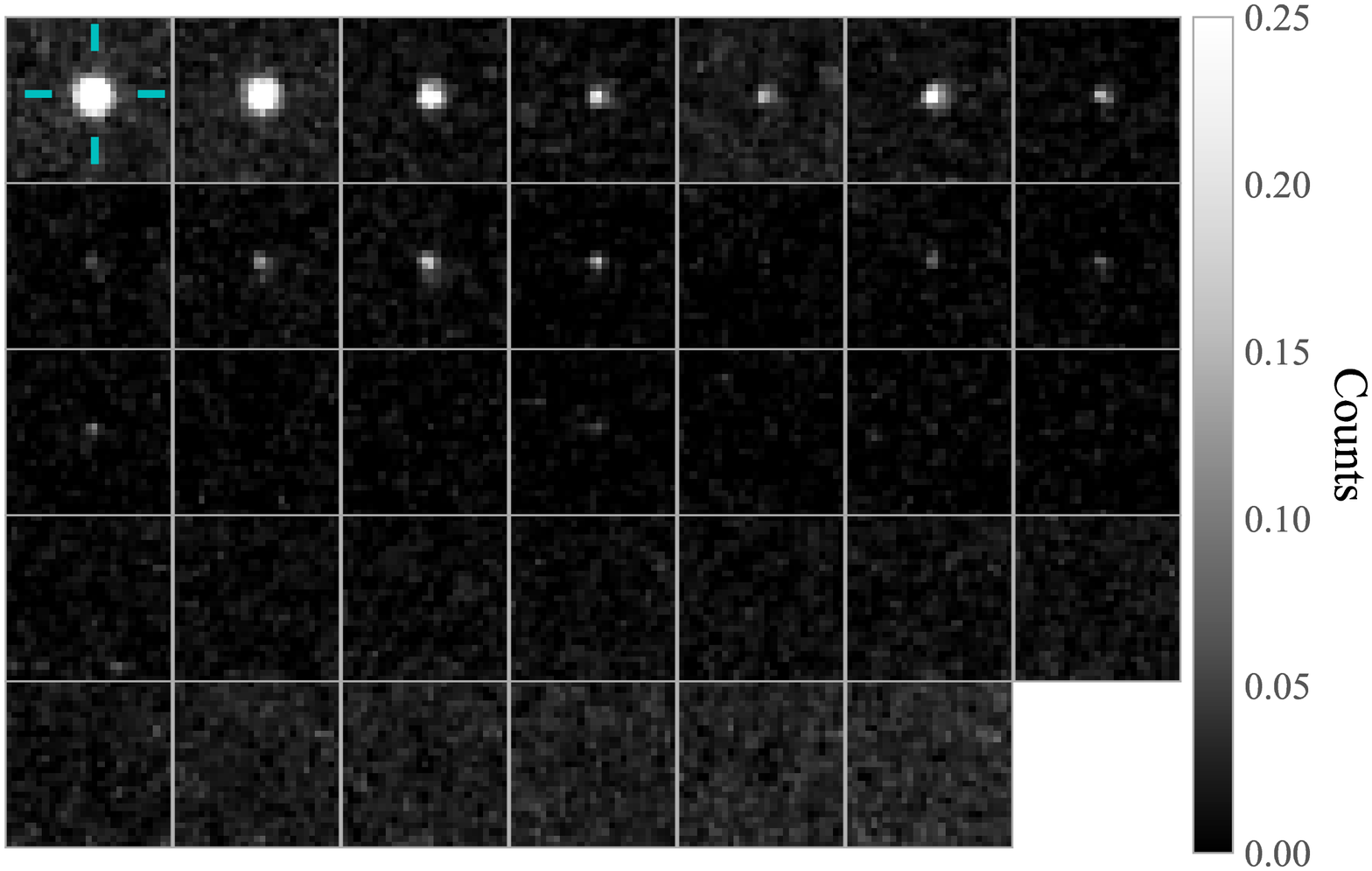}\label{fig:g}} 
  \hfill
  \subfloat[Nova 8]{\includegraphics[width=0.49\textheight]{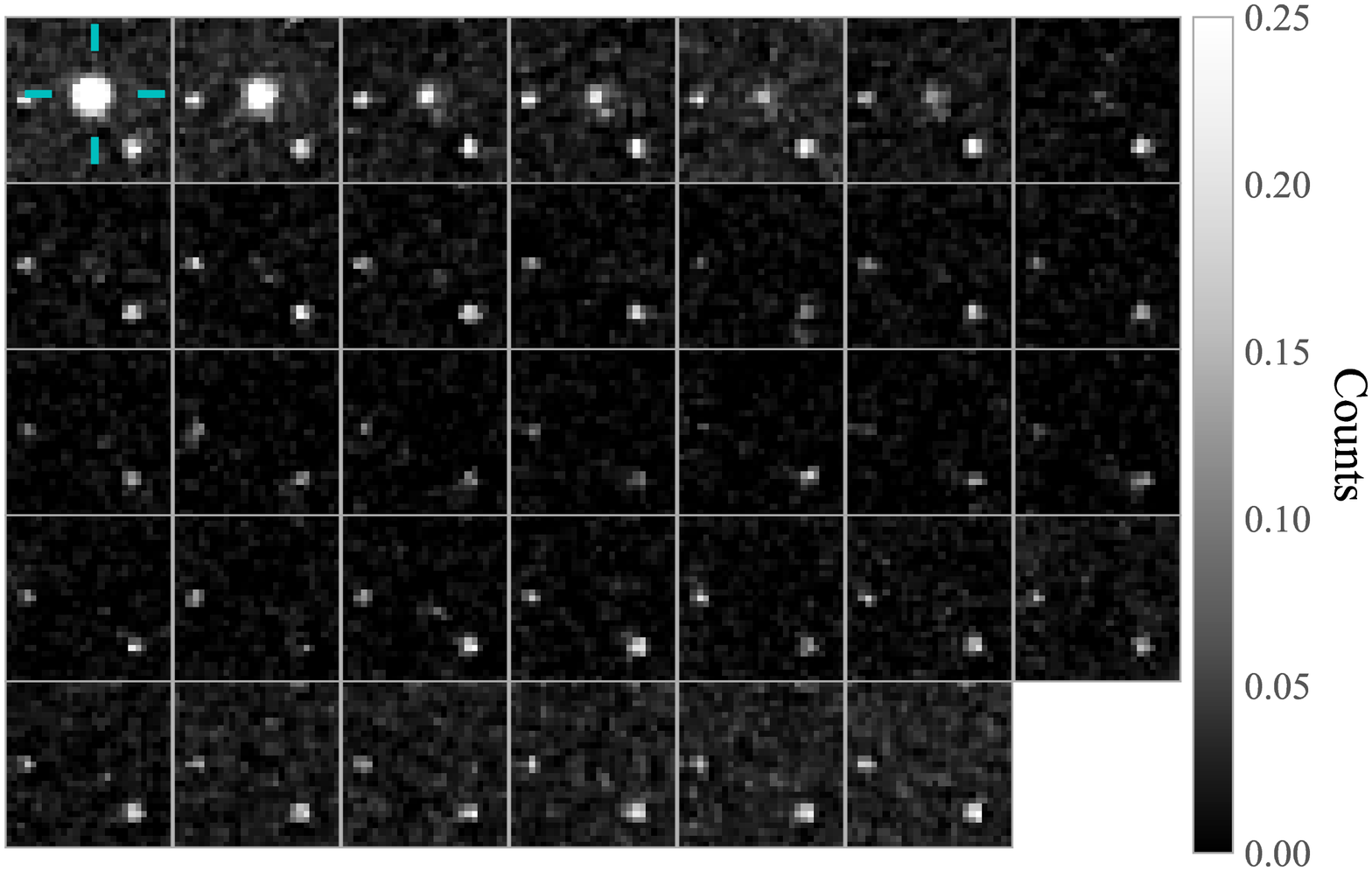}\label{fig:h}}
\end{center}
\end{sidewaysfigure*}

\begin{sidewaysfigure*}
\begin{center}
\centering
  \subfloat[Nova 9]{\includegraphics[width=0.49\textheight]{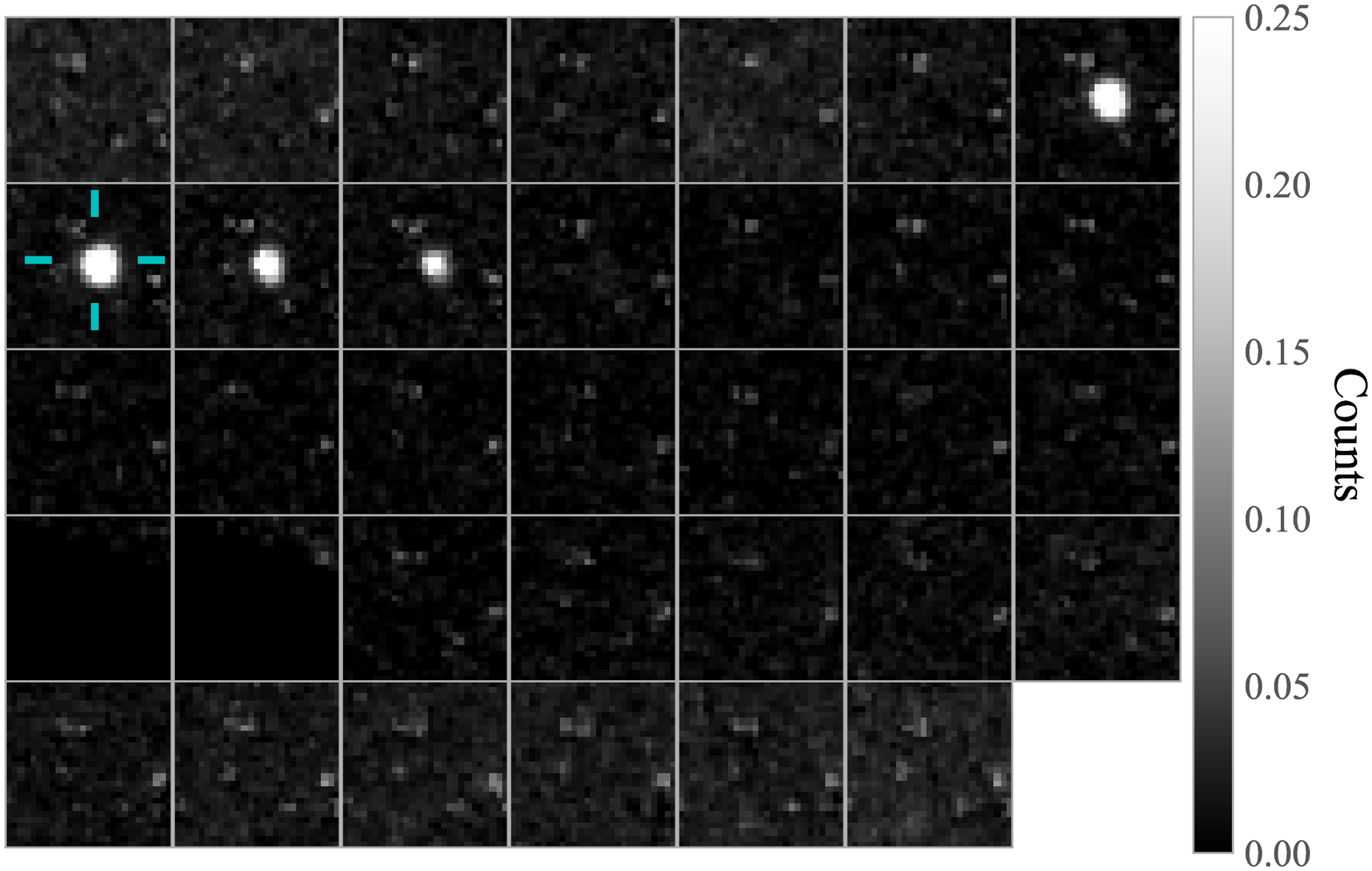}\label{fig:i}} 
  \hfill
  \subfloat[Nova 10]{\includegraphics[width=0.49\textheight]{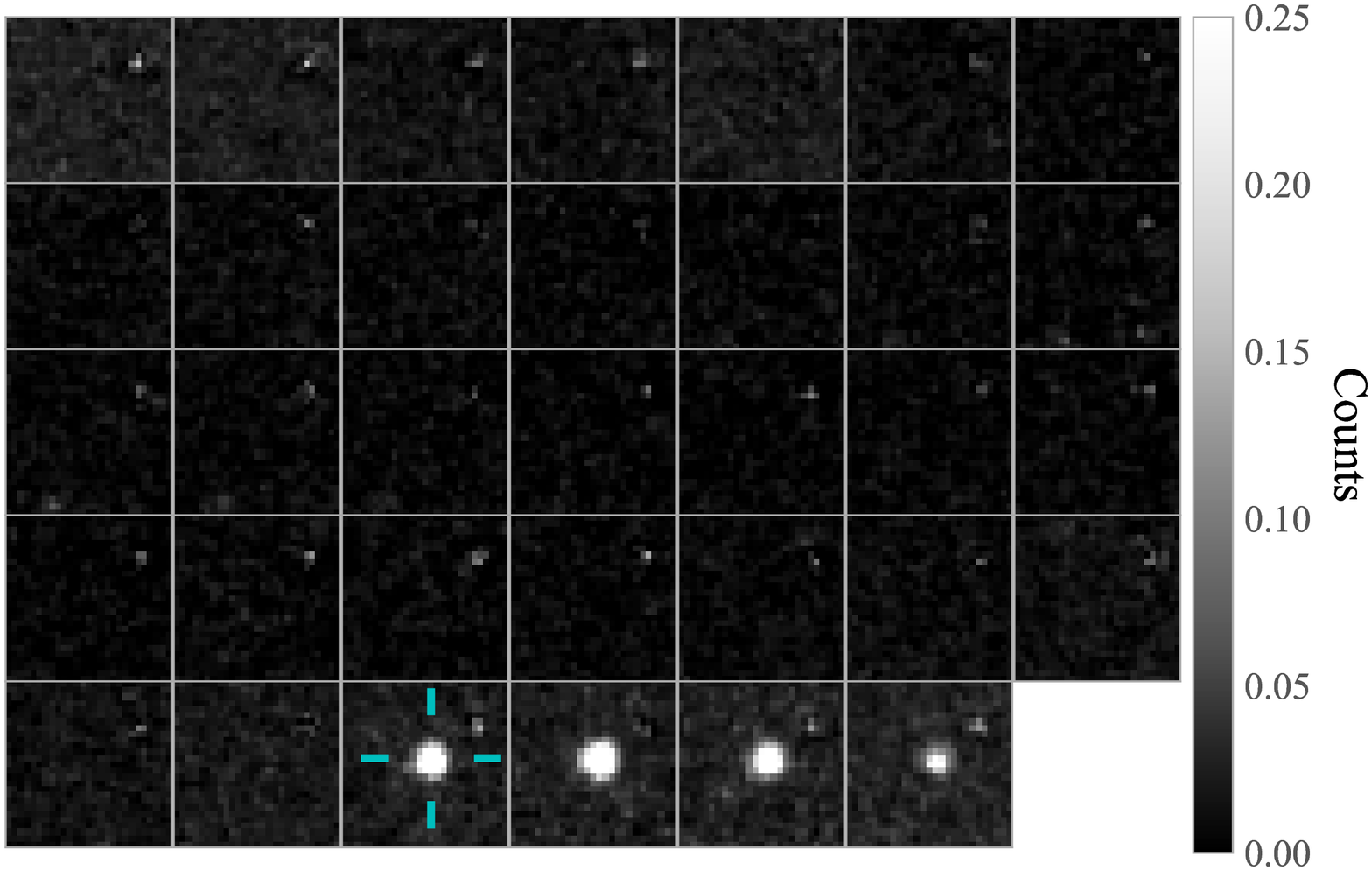}\label{fig:j}}
  \vfill
  \subfloat[Nova 11]{\includegraphics[width=0.49\textheight]{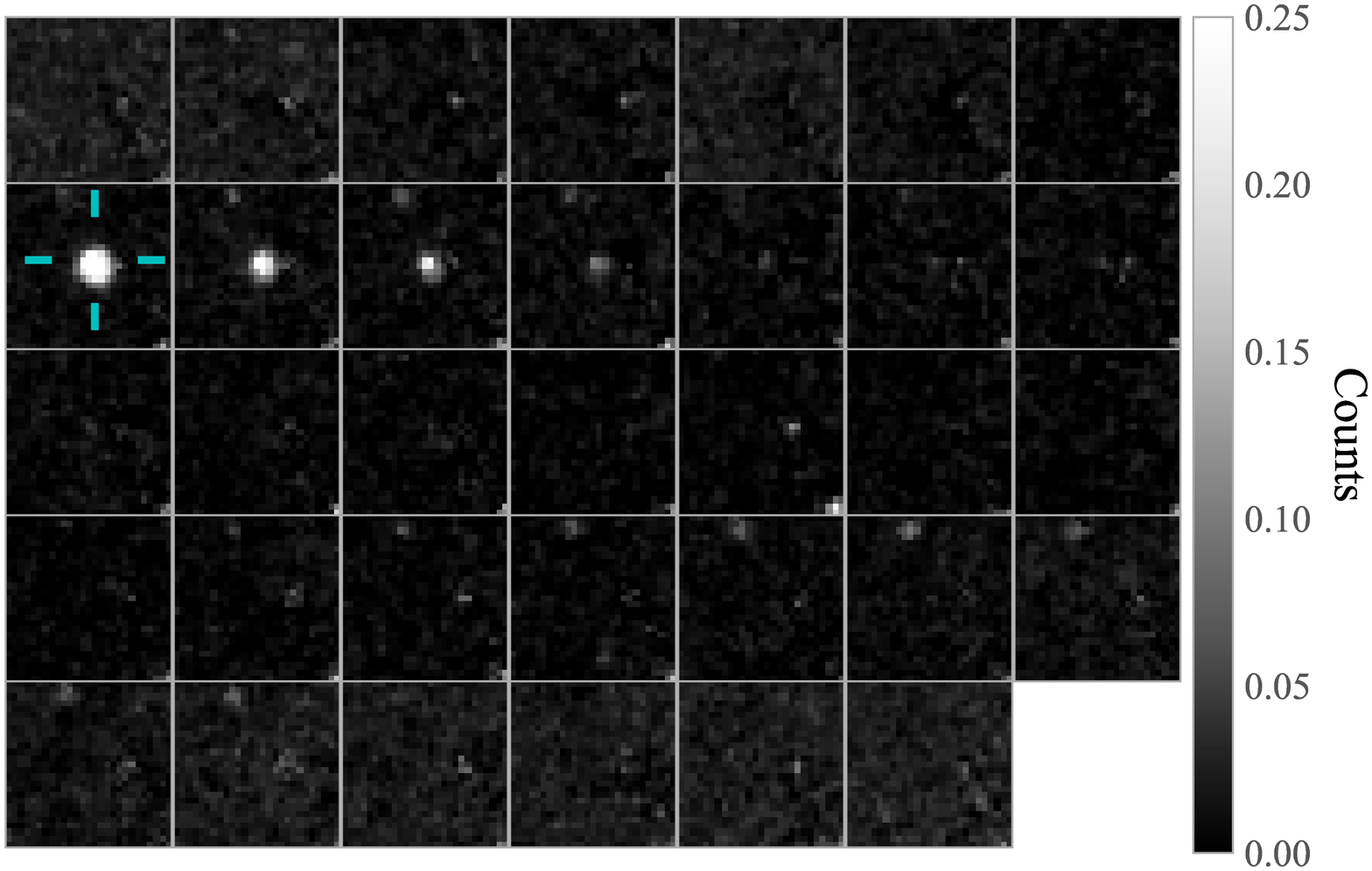}\label{fig:k}} 
  \hfill
  \subfloat[Nova 12]{\includegraphics[width=0.49\textheight]{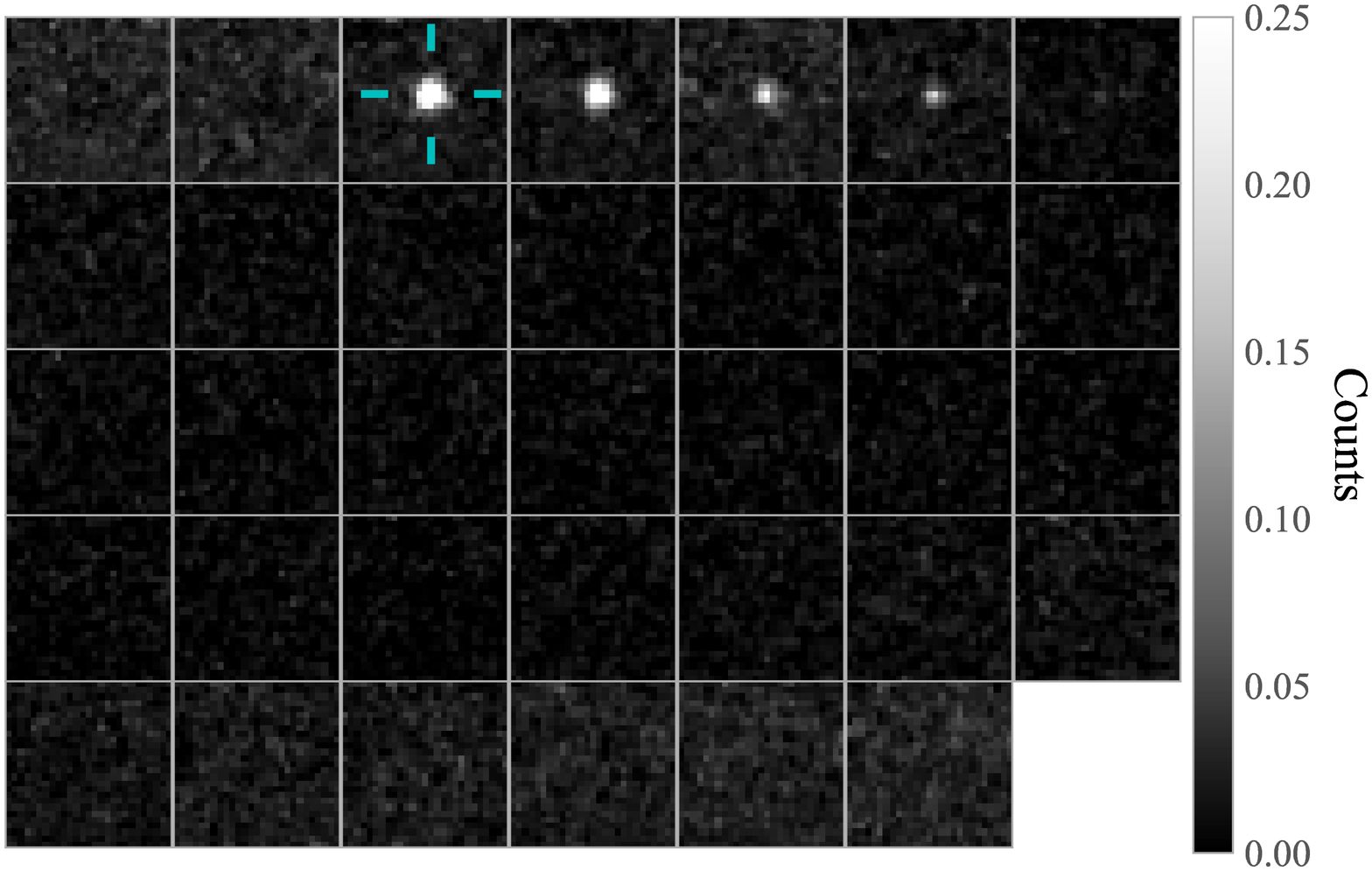}\label{fig:l}}
\end{center}
\end{sidewaysfigure*}

\bsp	
\label{lastpage}
\end{document}